\documentclass[superscriptaddress,showpacs,aps,twocolumn,prl,floatfix]{revtex4-2} %,linenumbers
\usepackage{eurosym}
\usepackage{amssymb}
\usepackage{amsmath}
\usepackage{amsfonts}
\usepackage{amsbsy}
\usepackage{graphicx}
\usepackage{bm}
\usepackage{xcolor}
\usepackage{hyperref}
\usepackage{units}
\usepackage{xspace}
\usepackage{verbatim}

\newcommand\BOg{$\mathrm{B}_{1g}$\xspace}
\newcommand\BTg{$\mathrm{B}_{2g}$\xspace}
\newcommand\sro{Sr$_{2}$RuO$_4$\xspace}

\def\Zopt{\mathcal{D}}

\begin{document}

\title{Signatures of Hund Metal and finite-frequency nesting in Sr$_2$RuO$_4$ Revealed by Electronic Raman Scattering} 

\author{Germ\'{a}n Blesio}
\email{blesio@ifir-conicet.gov.ar}
\affiliation{Jo\v{z}ef  Stefan  Institute,  Jamova  39,  SI-1000  Ljubljana,  Slovenia}
\affiliation{Instituto de F\'{\i}sica Rosario (CONICET) and Facultad de Ciencias Exactas, Ingenier\'{i}a y Agrimensura, 
Universidad Nacional de Rosario, 2000 Rosario, Argentina}

\author{Sophie Beck}
\affiliation{Center for Computational Quantum Physics, Flatiron Institute, New York 10010, USA}

\author{Olivier Gingras}
\affiliation{Center for Computational Quantum Physics, Flatiron Institute, New York 10010, USA}

\author{Antoine Georges}
\affiliation{Coll\`{e}ge de France, 11 place Marcelin Berthelot, 75005 Paris, France.}
\affiliation{Center for Computational Quantum Physics, Flatiron Institute, New York 10010, USA}
\affiliation{CPHT, CNRS, Ecole Polytechnique, Institut Polytechnique de Paris, Route de Saclay, 91128 Palaiseau, France.}
\affiliation{Department of Quantum Matter Physics, University of Geneva, 24 quai Ernest-Ansermet, 1211 Geneva, Switzerland}

\author{Jernej Mravlje}
\affiliation{Jo\v{z}ef  Stefan  Institute,  Jamova  39,  SI-1000  Ljubljana,  Slovenia}
\affiliation{Faculty of Mathematics and Physics, University of Ljubljana, Jadranska 19, 1000 Ljubljana, Slovenia}

\begin{abstract}
  We investigate the electronic Raman scattering of Sr$_2$RuO$_4$ using a material-realistic dynamical mean-field theory approach.   We identify the low-energy Fermi liquid behavior and point out that the enhanced Raman response at higher energies is a fingerprint of Hund metals. These signatures originate in the two-stage coherence of Hund metals and associated quasiparticle `unrenormalization'. In agreement with recent experimental observations, we find the \BOg and \BTg responses differ, but our calculations suggest a novel interpretation of this dichotomy. The \BOg response is dominated by the $xy$ orbital and the \BTg response 
  receives contributions from all orbitals and is strongly affected by previously unnoticed finite-frequency interband nesting. We calculate the vertex-corrections to Raman response and show that their effect is nonvanishing but small.
\end{abstract}

%\pacs{80.23.-b, 71.10.Hf, 75.20.Hr, 73.23.-b  }
 
\maketitle

\paragraph{Introduction.} 
Raman inelastic light scattering is a powerful probe of molecules and materials. 
Raman shifts and scattering intensities 
reveal vibrational modes, providing invaluable insights into molecules and compounds 
relevant to chemistry, biology, and materials science~\cite{schrader95}. 
Electrons in materials scatter light
inelastically too, and the associated electronic Raman
response~\cite{Devereaux2007} has been used to characterize the pseudogap
and superconducting phases of cuprates~\cite{LeTacon2006,Sacuto2011,Loret2019}, the nematicity in pnictides~\cite{Kretzschmar2016,Gallais2016, Yao2022} and to document strange-metal behavior and quantum criticality~\cite{Sen2020,Jost2022}, to name only a few.
An important advantage of Raman scattering is that by varying the
polarization of the incident and reflected light one can probe
different symmetry sectors and access the momentum dependence
of the scattering.  

Theoretically, Raman scattering has been
investigated with weak coupling approaches~\cite{Maiti2017,Udina2020} 
and, for strongly correlated models, with dynamical mean-field theory~\cite{Georges1996,Freericks2003,Shvaika2004,Shvaika2005,Devereaux2007,Medici2008,Lin2012} and determinantal quantum Monte-Carlo~\cite{Liu2021}.
Whereas phonon frequencies and associated Raman shifts are routinely
calculated~\cite{Baroni2001,Lazzeri2003}, electronic Raman scattering is rarely considered
and has been evaluated for a few cases only~\cite{Mazin2010,Valenzuela2013}. 
The rarity of material-realistic approaches represents a significant hindrance to the interpretation of experiments, highlighting the need for further work in this area. Our paper develops such an approach and
applies it to a case of great current interest, 
the layered ruthenate compound Sr$_2$RuO$_4$.

The motivation for our investigation is Ref.~\citenum{Philippe2021} that recently reported 
experimental measurements of the electronic Raman response in Sr$_2$RuO$_4$. 
The much discussed Sr$_2$RuO$_4$ compound is a multi-orbital layered 
unconventional superconductor~\cite{Mackenzie2003} displaying Fermi liquid behavior 
below a low quasiparticle coherence scale 
(e.g. resistivity $\propto T^2$ below $\sim 25\,$K).  On a broader energy and temperature range, \sro has been 
characterized~\cite{Mravlje2011,Mravlje2016,Kugler2020} as a Hund metal~\cite{Yin2011,Georges2013}. 
A hallmark of Hund metals is that local orbital fluctuations become 
quenched at a distinctly higher energy scale than local spin fluctuations 
~\cite{Yin2012,Aron2015,Stadler2015,Mravlje2016,Horvat2016}.
This two-stage screening leads to a distinctive non-Fermi liquid feature at a 
characteristic intermediate energy scale in the one-particle spectral function and self-energy~\cite{Wadati2014,Stricker2014,Stadler2015,Kugler2019}. 
Another hallmark of Hund metals is a tendency towards orbital differentiation, 
manifested by distinct values of quasiparticle renormalizations in bands 
spanned by different orbitals. 
In Sr$_2$RuO$_4$ the broader quasi-two-dimensional $\gamma$-band, predominantly spanned by the $xy$-orbital 
and displaying a van Hove singularity close to Fermi level,
is more renormalized (i.e., it has a higher mass enhancement $\sim 5-6$ compared to band theory) ~\cite{Mackenzie2003,Mravlje2011} than the $\alpha$/$\beta$ quasi-one-dimensional sheets spanned by the $xz/yz$ orbitals (mass enhancement $\sim 3.5$).

In Ref.~\citenum{Philippe2021}, they measured the Raman spectra 
in different channels, found a less coherent response in  the B$_{2g}$ channel, and interpreted the dichotomy between the B$_{1 g}$ and the B$_{2g}$ response in terms of the orbital effects, referring to the effect as the `orbital-dichotomy'.  
This interpretation may appear natural since a simple approximation 
$\sim \partial^2\epsilon (\mathbf{k})/\partial k_x  \partial k_y$ 
to the Raman vertex in this channel suggests that it is insensitive to the 
more coherent quasi-one-dimensional $\alpha/\beta$ bands, leading to the interpretation~\cite{Philippe2021} 
that the B$_{2g}$ response may originate dominantly from the $xy$ orbital and the 
B$_{1g}$ channel from the $xz/yz$ ones. 
However, the quasiparticle response reported for the B$_{2g}$ channel 
is quite faint even at low$-T$~\cite{Philippe2021}, which seems at odds with 
Fermi liquid behavior and raises questions about the appropriate interpretation. 
What are the signatures associated with the Fermi liquid regime and those associated 
with the  higher energy non-Fermi liquid Hund metal effects? 
How does the orbital differentiation manifest itself in the Raman response?
These questions can only be convincingly addressed theoretically in a materials-realistic 
approach.
  
Here, we answer these questions by calculating and analyzing 
the electronic Raman response  of Sr$_2$RuO$_4$ in a 
dynamical mean-field theory (DMFT) framework based on first-principles 
density-functional theory (DFT) electronic structure. 
The Raman vertices are calculated in a matrix-valued effective mass approximation. By comparison with simplified model calculations we disentangle the Fermi liquid response from the higher-energy non-Fermi liquid response and show that the latter 
reveals characteristic spectral features of Hund metals.
We also investigate why the Raman responses in the \BOg and \BTg channels differ. 
We show that, in contrast to the interpretation of Ref.~\citenum{Philippe2021}, the 
\BOg channel is dominated by the $xy$ orbital and find that the 
\BTg response displays a spectral feature associated with the enhancement of  
inter-band contributions by an interesting finite-frequency nesting property.

\paragraph{Methods.}
We perform DFT+DMFT calculations
for a DFT-derived Wannier Hamiltonian~\cite{Wannier90} spanning the t$_{2g}$ 
bands of \sro as described in Ref.~\citenum{Tamai2019}, 
taking into account the correlation-enhanced spin-orbit coupling 
$\sim 0.2\,$eV~\cite{Kim2018,Zhang2016}. 
We use the hybridization-expansion quantum Monte-Carlo solver 
from the TRIQS software library 
~\cite{triqs,triqs_wien2k_interface,triqs_wien2k_full_charge_SC,TRIQS/DFTTools}, using the interaction parameters $U=2.3\,$eV and $J=0.4\,$eV, which were originally estimated from constrained-RPA calculations~\cite{Mravlje2011} and shown to give results consistent with several experimental observables~\cite{Mravlje2011,Mravlje2016,Sutter2017,Strand2019,Tamai2019,Kugler2020,Suzuki2022,Hunter2023}.
The Matsubara self-energies are continued to real-frequency using Pad\'e interpolation. We checked that the results do not change significantly if we use Maximum Entropy analytical continuation.
The Raman response is given by (in units where $e^2=1$ and $\hbar=1$):
\begin{align}
  \chi''_\mu(\Omega) = \pi \sum_\mathbf{k} \int d \omega & \left\{  \left[f(\omega)-f(\omega+\Omega)\right] \right.     \label{eq:transp_dist}    \\
     &\left. \times \,  \mathrm{Tr}  \, \gamma^\mu_\mathbf{k} \, {\cal A}_\mathbf{k}(\omega) \, \gamma^\mu_\mathbf{k} \, {\cal A}_\mathbf{k}(\omega + \Omega) \right\}. \notag
\end{align}
In this expression, both the Raman vertex $\gamma^\mu_\mathbf{k}$ in channel $\mu$ and spectral functions ${\cal A}_\mathbf{k}$ at momentum $\mathbf{k}$  are $6\times6$ matrices in spin and orbital space; $f(\omega$) is the Fermi function. 
 In Eq.~\ref{eq:transp_dist} we retained only the so-called bubble contributions to the response~\cite{Mahan2000}. It is well known that for the case of optical conductivity the vertex corrections vanish in DMFT due to the oddness of 
the velocity matrix elements upon inversion for centrosymmetric systems~\cite{Khurana1990,Georges1996}.
This argument does not hold however for the Raman response, because the Raman matrix elements transform differently. 
In SM 8 we show that (i) when the interband terms are neglected, the vertex corrections in the considered channels vanish by reflection symmetries, and (ii) numerically test that their contribution remains small even when the interband terms are included. 
Following Ref.~\citenum{Valenzuela2013}, we calculate the Raman vertices in a matrix valued effective mass approximation generalizing the approach of Ref.~\citenum{Shastry1991}:  
\begin{align}
&\left[\gamma^\mu_\mathbf{k}\right]_{\nu\nu'}= \sum\limits_{\alpha\beta} e^i_\alpha \frac{\partial^2 H^{(W)}_{\nu\nu'} ~}{\partial k_\alpha \partial k_\beta} e^s_\beta \quad \nu,\nu'\text{ band indices}
\label{eq:raman_vertex} \\
&\gamma^{B_{1g}}_\mathbf{k}=\frac{1}{2}\left( \frac{\partial^2 H^{(W)}}{\partial k_x^2 \hfill} - \frac{\partial^2 H^{(W)}}{\partial k_y^2 \hfill}\right) ~ ; ~ \gamma^{B_{2g}}_\mathbf{k}=\frac{\partial^2 H^{(W)}}{\partial k_x \partial k_y}\notag 
\end{align}
expressed in terms of the second derivative of the Wannier Hamiltonian $H^{(W)}$ with respect to $\mathbf{k}$ (in a direction related to the polarization of the incoming/scattered light $e^i$ and $e^s$). 
We emphasize that, importantly, this expression includes intraband and interband contributions. The derivatives were performed as outlined in Ref.~\citenum{Yates2007} using the Wannier interpolation as implemented in the WannierBerri library~\cite{Destraz2020,Tsirkin2021}. We use a $60\times 60 \times 60$ Monkhorst-Pack grid to sample the Brillouin zone, which we verified to be sufficient for convergence. 

\begin{figure}[b]
\begin{center}
\includegraphics*[width=0.9\columnwidth]{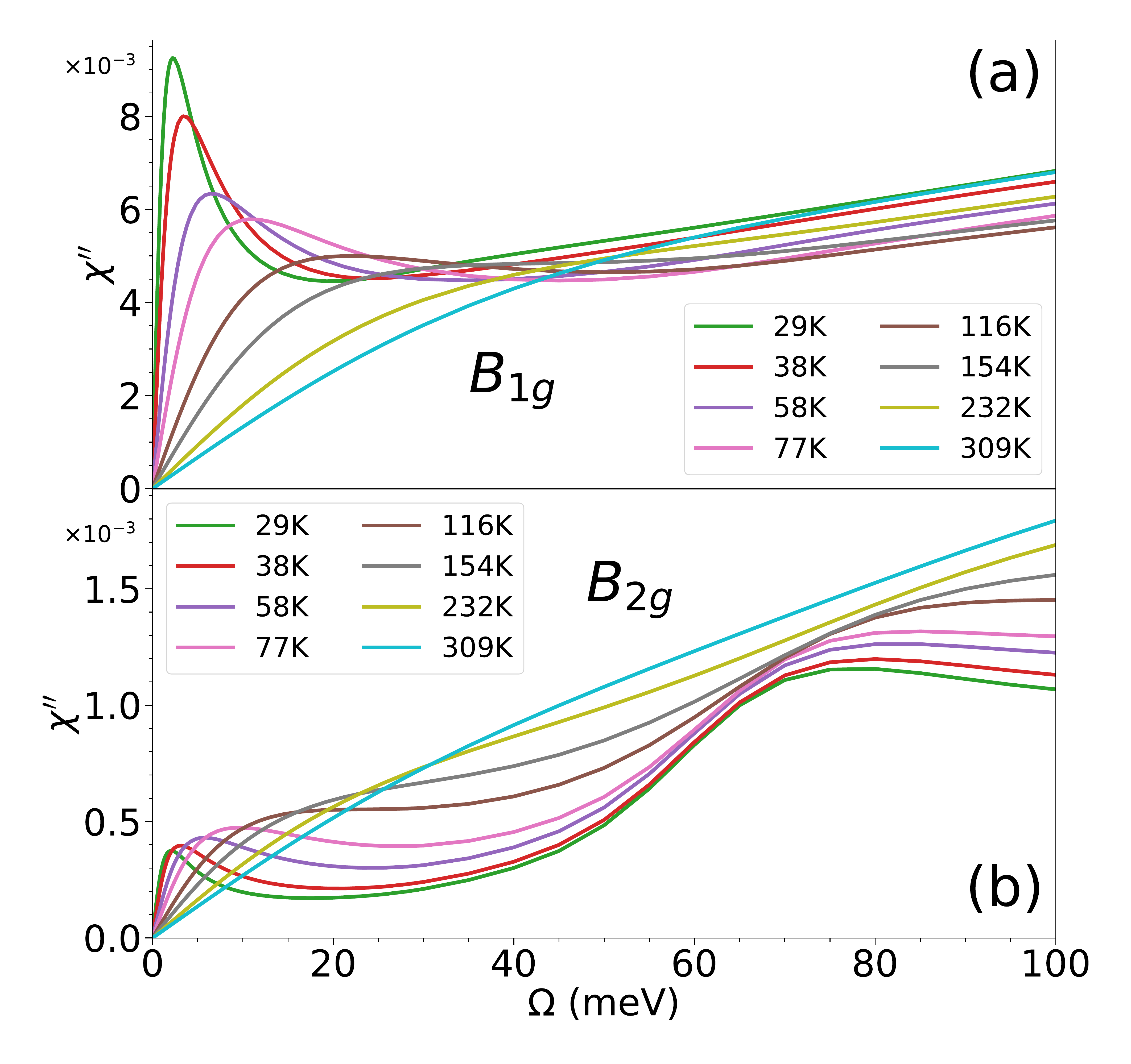}
\caption{Raman response in B$_{1g}$ (top panel) and B$_{2g}$ (bottom panel) channels for different temperatures. The susceptibilities are expressed in atomic units (i.e. setting the prefactor $e^2/(\hbar a_0 m_e)$ equal to unity).
}
\label{fig:raman_general}
\end{center}
\end{figure}

\paragraph{Results.} 
Figure~\ref{fig:raman_general} displays the temperature ($T$) and frequency dependence of the calculated Raman responses. The top panel depicts the results in the \BOg channel. For low-$T$, one sees a low-frequency Drude peak that is followed by a moderate increase at a higher frequency, a dependence that is characteristic of the Fermi liquid coherent response~\cite{Freericks2001,Philippe2021,Berthod2013}. This dependence persists up to $T\sim200\,$K. For higher $T$ the peak-dip-increase structure is lost and one observes only a monotonous increase with frequency in the presented frequency range.

In the \BTg channel shown in Fig.~\ref{fig:raman_general}~(b), one also sees a Drude peak but its intensity is smaller compared to an additional feature that takes the shape of a peak centered around $80\,$meV at low-$T$. The intermediate minimum disappears at a somewhat lower $T$ than observed in the \BOg. At $T\sim 100-150\,$K in \BTg, the two-peak structure transforms  to that of a plateau followed by a steeper increase, whereas in \BOg this growth of the response above $40\,$meV remains mild.
Qualitatively similar features, also with a smaller overall intensity but with a steeper increase with frequency in the  \BTg case, was also observed in experiments~\cite{Philippe2021}.

\begin{figure}[h]
\begin{center}
\includegraphics*[width=0.9\columnwidth]{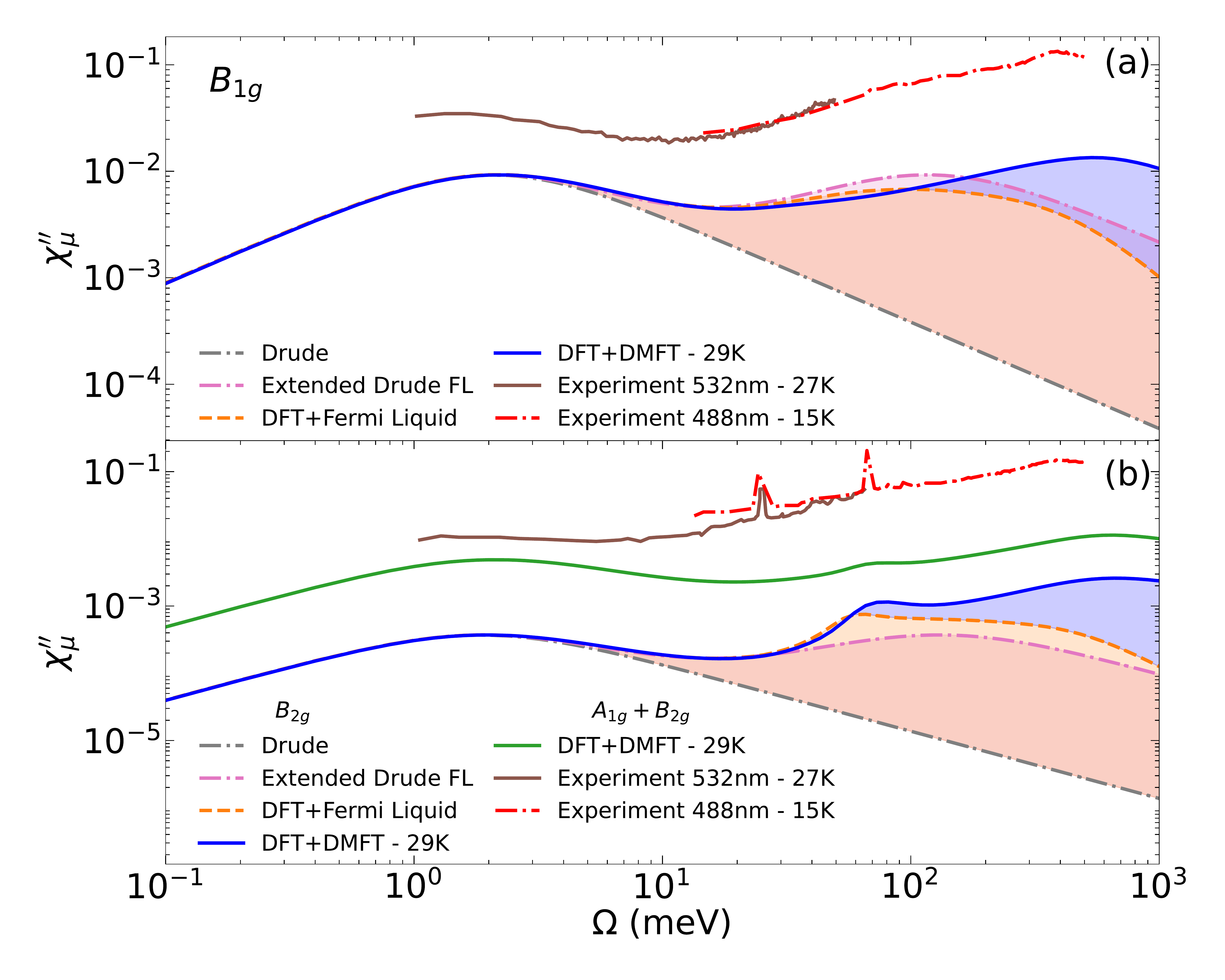}
\caption{
Raman response for the B$_{1g}$ (a) and B$_{2g}$ (b) channel. 
In addition to the full DFT+DMFT result at $T=29\,$K (solid blue line), we also display 
a calculation where the DMFT self-energy for each orbital is replaced by a Fermi liquid expression 
consistent with its low-energy behavior (orange dashed). 
The results are compared to: 
a simple Drude fit with an optical scattering rate $\Gamma^*=\mathrm{const.}$ (gray dashed-dotted) 
and an extended Drude fit with a Fermi liquid optical scattering rate 
$\Gamma^*  \propto \Omega^2 + (2 \pi k_B T)^2$ (pink dashed-dotted). 
We also display the experimental results from Refs~\citenum{Philippe2021} and \citenum{philippe21phd} 
(shifted away vertically for clarity).
On the bottom panel, the same curves are plotted for \BTg, except for the experimental data which are 
currently available in a broader frequency range only for the A$_{1g}+$B$_{2g}$ channel~\cite{Philippe2021,philippe21phd}. 
For direct comparison,  we also plot the DFT+DMFT results for that channel (green). 
See Supplemental Material S7 for the expression of the Raman vertex in the A$_{1g}+$B$_{2g}$ channel.}
\label{fig:FL_NFL}
\end{center}
\end{figure}

In Fig.~\ref{fig:FL_NFL}, we display the calculated low temperature $T=29\,$K 
spectra in a broader energy range. To gain insight into the different features 
of the spectra, we also display  
(i) a simple low-energy Drude fit $\chi'' \propto \omega /(\omega^2+{\Gamma^*}^2)$ with constant scattering rate $\Gamma^*$ 
(ii) an extended Drude fit using a frequency-dependent 
Fermi liquid scattering rate  $\Gamma^*(\omega) \propto [\omega^2 + (2\pi k_B T)^2]$
and (iii) a `DFT+Fermi Liquid' approximation in which 
the DMFT self-energies have been replaced 
by their low-energy Fermi liquid form for each orbital: 
$\Sigma_m(\omega) = (1-1/Z_m)\omega -i A_m (\omega^2 + \pi^2 k_B^2 T^2)$
in the full expression (\ref{eq:transp_dist}). 
The calculated spectra deviate from the simple Drude fit above $\sim 3\,$meV in both channels. 
A better description of the data up to $\sim 20-30\,$meV is obtained using 
the generalized Drude fit, which  
on a log-log scale has the shape of two peaks of similar height separated by a dip. 
This dip signals the crossover from the regime in which the scattering rate 
is dominated by temperature to that in which it is dominated by its 
Fermi liquid frequency dependence~\cite{Berthod2013,Stricker2014}.  
It is apparent in the DFT+DMFT spectra around $25\,$meV and also (at a somewhat lower 
frequency) in the \BOg experimental spectra, as noted in Ref.~\citenum{Philippe2021}. 

In the \BOg channel, deviations between the calculated DFT+DMFT spectra and Fermi liquid behavior are apparent on Fig.~\ref{fig:FL_NFL}~(a). 
Weak deviations are first noticeable at frequencies of 
$\sim 20-30\,$meV, 
where the real-part of the self-energy (see Supplemental Material S2) deviates from the strict linear-in-$\omega$ Fermi liquid behavior (which also causes `kinks'  observed in angular resolved photo-emission experiments at similar frequencies~\cite{Aiura2004,Ingle2005,Tamai2019}). As the frequency increases, the deviations become much more pronounced above $\hbar\omega\sim 100\,$meV, where the DFT+Fermi liquid result starts to drop, while the full DMFT result continues to increase. 
The enhancement of the Raman response in this energy range is linked to 
a non-Fermi liquid feature of the DMFT self-energies $\Sigma$, which is a hallmark of Hund metal behavior. 
The real part $\mathrm{Re}\Sigma$ for both orbitals displays a minimum and 
an abrupt change of slope for $\hbar\omega\gtrsim 100\,$meV~\cite{Mravlje2011} (see Supplemental Material S4), 
leading to an `unrenormalisation' 
and waterfall in the quasiparticle dispersion. This feature that also causes a shoulder structure in the quasiparticle peak~\cite{Wadati2014} has been shown to enhance the optical conductivity in this frequency range~\cite{Stricker2014}. We note that in Ref.~\citenum{Stricker2014} the relation to Hund physics was missed. 
In Supplemental Material S4, we compare the results also to a calculation at increased occupancy $N=5$, explicitly showing that the enhanced Raman response in the $100-500\,$meV range is due to the Hund metal inner structure of the quasiparticle peak. 
This effect is due to the two-stage 
coherence which is a hallmark of Hund metals: distinct coherence scales for spin and orbital 
degrees of freedom lead to additional structures in both one- and two-particle responses~\cite{Stadler2015,Mravlje2016,Horvat2019,Wang2020,Walter2020}.

Figure~\ref{fig:FL_NFL} also displays the experimental data 
from Refs~\citenum{Philippe2021} and \citenum{philippe21phd}.
Our analysis suggests
that the observed increase of the experimental spectra, peaking at $\sim 400\,$meV, is a manifestation 
of the Hund metal non-Fermi liquid feature discussed above. 
Comparison to the DFT+Fermi liquid approximation demonstrates that this increase cannot be interpreted in terms of inter-band effects, which are fully taken into account in this approximation. 
Furthermore, in Refs~\citenum{Philippe2021} and \citenum{philippe21phd} it was shown that 
luminescence cannot account entirely for this increase. 
The clear signature of Hund metal behavior in Raman spectroscopy 
is the first main result of our work.

\begin{figure*}[t]
\begin{center}
\includegraphics*[width=1.9\columnwidth]{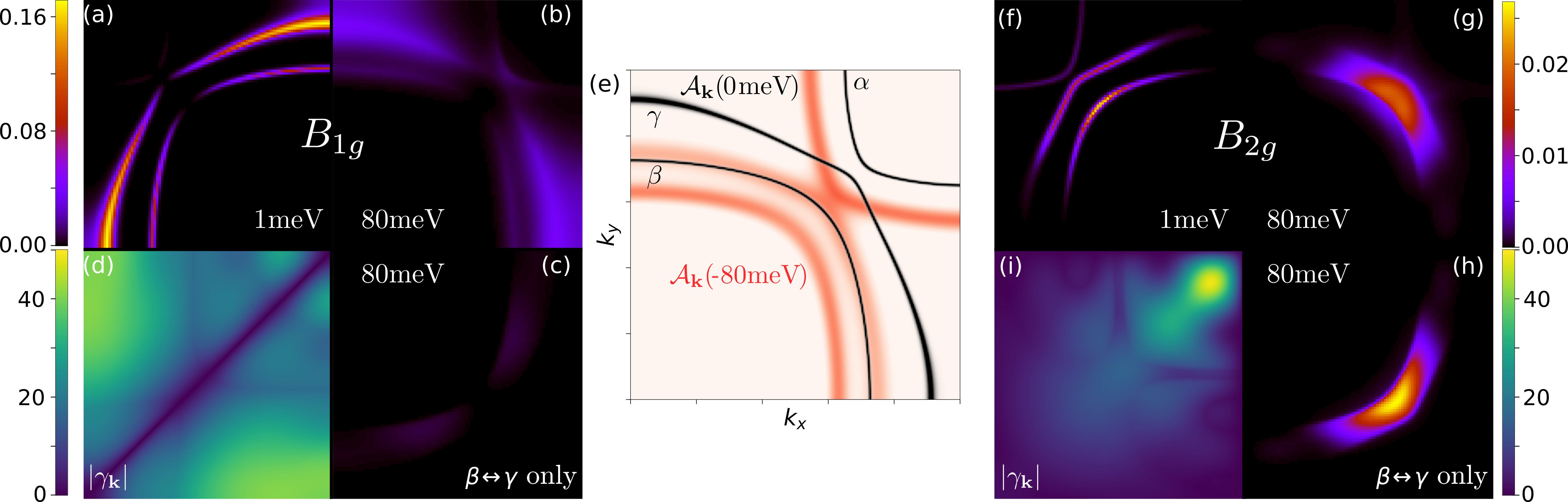}
\caption{Momentum-resolved Raman response. 
  The $(k_x,k_y)$ map of contributions to  total $\chi_{B_{1g}}''$ at $\Omega=1$ meV (a) and  at $\Omega=80$ meV (b). (c) The interband  $(\beta \leftrightarrow \gamma$) contributions   at $\Omega=80$ meV. (d) The magnitude of the Raman vertex $|\gamma^{B_{1g}}|  = \sum_\nu |\gamma^{B_{1g}}_{\nu \nu}|$. Panel (e) depicts the spectral function at two energies.   In (f-h) the same information as in (a-d) is shown but for the B$_{2g}$ channel. The response and vertex values are all in atomic units.}
\label{fig:nesting_vertex}
\end{center}
\end{figure*}

Figure~\ref{fig:FL_NFL}~(b) shows the results for the \BTg channel, along with the A$_{1g}$ + B$_{2g}$ channel for which experimental data are available in a wider frequency range~\cite{philippe21phd}. 
The results are similar to those in the \BOg channel  
and notably also display the non-Fermi liquid increase above $100\,$meV. 
In addition, the spectra display a clear `hump' around $80\,$meV. 
This hump is visible in A$_{1g}$ + B$_{2g}$ and is even more pronounced 
in \BTg but does not appear in the \BOg channel. 
As the hump also appears in the DFT+Fermi liquid approximation, it cannot be attributed to non-Fermi liquid physics. Moreover, if one isolates the response corresponding to either of the two orbitals individually, one does not find a hump at that frequency (see Supplemental Material S3). Instead, we find the hump is a consequence of unexpectedly strong interband transitions.

\begin{figure}[h]
\begin{center}
\includegraphics*[width=0.9\columnwidth]{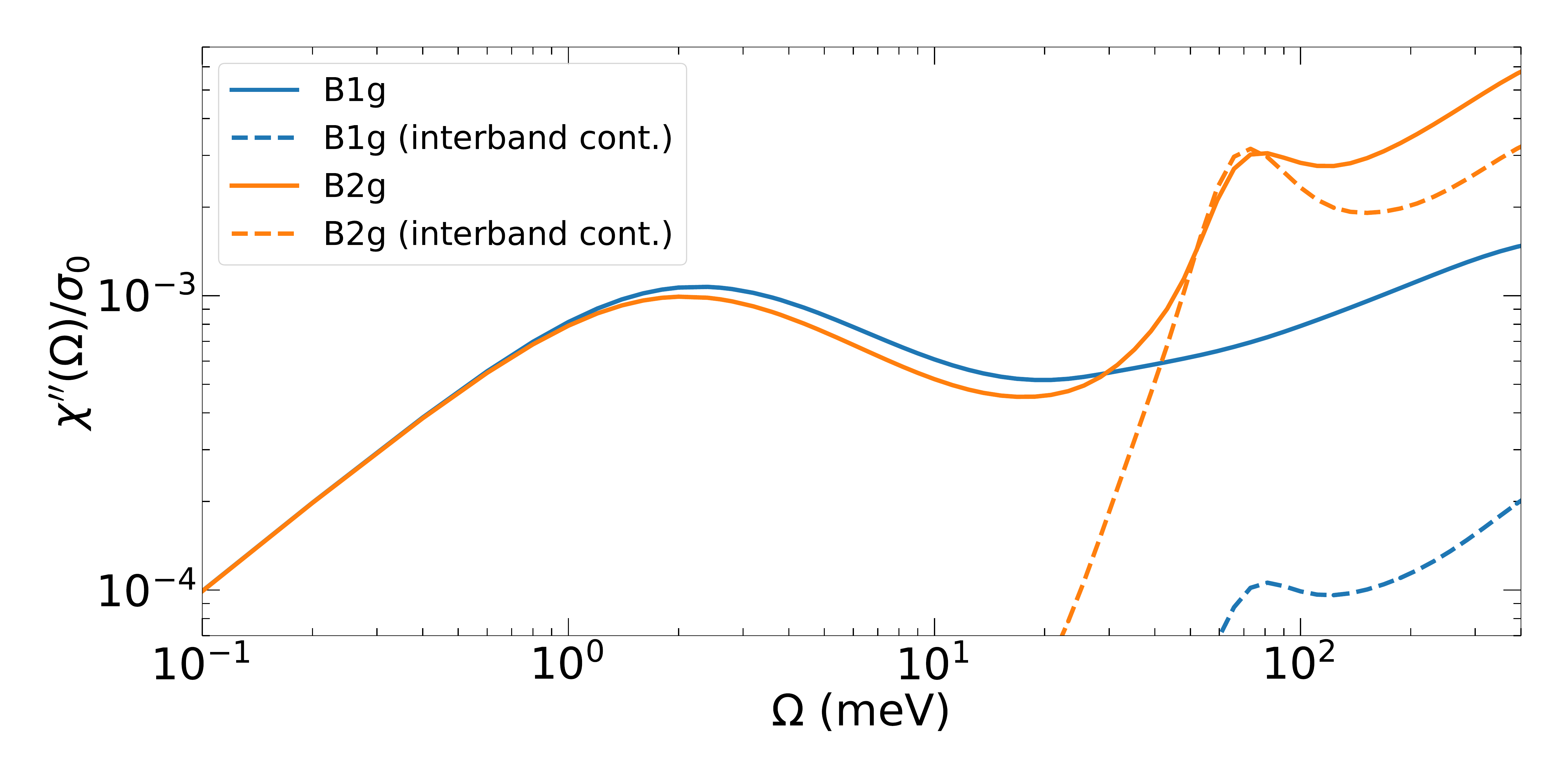}
\caption{
  Normalized Raman response for  B$_{1g}$ (blue) and B$_{2g}$ (orange) at $29\,$K. The dashed lines indicate the interband contributions (see Supplemental Material S6).
}
\label{fig:allen}
\end{center}
\end{figure}

We demonstrate this on Fig.~\ref{fig:allen}, 
where we plot the full DFT+DMFT results next to a calculation that retains only the interband contributions to the response. We observe that whereas in \BOg the interband terms represent a small part of the full response, 
their contribution becomes dominant in \BTg around $80\,$meV. 
Additionally, we observe that at low energies, the data in \BOg and \BTg behave similarly. 
This raises the
question of identifying which orbitals are probed by each
channel and whether a particular channel acts as an ‘orbital
filter'.
The remainder of the paper discusses this point and explains why the interband contribution at $80\,$meV is so strong.  

In Fig.~\ref{fig:nesting_vertex}~(a,f), we present a map of the contribution of different momenta to $\chi''$, 
i.e. of the momentum-resolved Raman response corresponding to the integrand in Eq.~(\ref{eq:transp_dist}), 
at a low frequency ($1\,$meV) for the \BOg and \BTg cases, respectively. This map exhibits the expected 
symmetry with maxima along the zone verticals/horizontals for \BOg and along 
the zone diagonals for \BTg. 
This is caused by the momentum dependence of the corresponding vertices - see panels (d) and (i) depicting their magnitude. 
On panels (a,f), one can distinguish the individual Fermi surface sheets and see that dominantly $\gamma$ and $\beta$ sheets contribute to both responses and that the \BOg response is primarily associated with the $\gamma$ sheet, which is composed primarily of the $xy$ orbital along the directions~\cite{Tamai2019} where the \BOg vertex is large (see also Supplemental Material S5 for discussion of orbital character of different channels). 
More precisely, the  weights of the \BOg signal on sheets $\alpha$, $\beta$, $\gamma$ are respectively 0\%, 16\%, 84\% 
while they are are 5\%, 44\% and 51\% for the \BTg signal. 
The $\alpha$ sheet contributes minimally to both responses due to the low magnitude of the vertices 
in the corresponding part of the Brillouin zone. 
From this analysis, it is clear that the \BOg response is dominated by the $xy$ orbital 
(with some smaller contribution from $xz/yz$), while 
in contrast the \BTg channel does not filter out a specific orbital contribution.
 
Panels (b) and (g) depict the Raman response at $80\,$meV. In the \BOg channel, the momentum map is essentially a broadened version of the low-frequency case and one can still resolve the individual sheets. Conversely, in the \BTg channel, the momentum structure changes, revealing a strong broad feature appearing between the $\beta$ and $\gamma$ sheets. Panels (c) and (h) again present a map of the response at $80\,$meV but retaining only the contributions that involve  interband $\beta$-$\gamma$ transitions. This clearly shows that the \BTg signal in panel (g) is dominated by those transitions, whereas for \BOg (c) their contribution is weak.
  
Why do the interband transitions become important only around $80\,$meV?  
Panel (e) shows an overlay of the momentum resolved spectral function evaluated at $\omega=0$  (black, corresponding to the Fermi surface) and at $\omega=-80\,$meV (red, corresponding to the renormalized equal-energy surface for quasiparticles at that binding energy). It is evident that the two spectral functions nearly completely nest, leading to strong interband effects whenever the vertex allows for such transitions. 
This motivates us to return to revisit the \BOg case in panel (c). 
We find that actually the interband contributions are of comparable magnitude in the two-channels 
(notice the different scales), but in the \BOg the magnitude of the intraband transitions is higher, resulting in less prominent interband contributions in the \BOg response (see also Supplemental Material S6).

The interband transitions also affect interpretation of the extended Drude modelling of the data. In Supplemental Material S1 we demonstrate that consistently with Ref.~\cite{Philippe2021} we find larger  mass enhancements in \BTg channel, but we stress their large magnitude is due to the interband-transitions.

\paragraph{Conclusions.} 
In summary, we used a material-realistic DMFT approach with Raman vertices evaluated in a matrix-valued effective mass approximation to calculate the Raman response in \sro.  
We documented the Fermi liquid regime at low energy and argued
that the enhanced response at higher energies is a consequence of the inner structure of quasiparticle peak, a characteristic fingerprint of Hund metals. 
We found that the \BTg response is strongly influenced by interband contributions 
that are enhanced by a finite-frequency nesting effect. 
Whether this effect leads to a possible feedback of the corresponding fluctuations into other physical 
properties and is visible 
in other spectroscopies~\footnote{ARPES spectra at $\sim 80\,$meV binding energy appear consistent with this nesting property (A.Tamai, private communication).}
 are interesting questions for future investigations. 
 Our results emphasize that this rarely discussed effect~\cite{Youn2004} is to be considered in other materials as well.  
On the methodology side, our investigation highlights the importance of material-realistic calculations of the Raman response and the approach developed in this work opens up possibilities for investigations 
of other materials with strong electronic correlations.

\textit{Acknowledgments.}
We thank Yann Gallais and Jean-C\^{o}me Philippe for very useful discussions and for sharing their raw data 
and also acknowledge discussions with A.~Tamai, F.~Baumberger and A.~Hunter. 
GGB and JM are supported by Slovenian Research Agency (ARRS) under Grant no. P1-0044 and J1-2458. 
The Flatiron Institute is a division of the Simons Foundation.

\providecommand{\noopsort}[1]{}\providecommand{\singleletter}[1]{#1}%

\onecolumngrid
\newpage

\begin{center}

{\large\textbf{\boldmath
Supplemental Material\\ [0.5em] {\small to} \\ [0.5em]                                                                                                                                                                            
Signatures of Hund Metal and finite-frequency nesting in Sr$_2$RuO$_4$  
Revealed by Electronic Raman Scattering}}\\[1.5em]

Germán Blesio,$^{1,2,*}$ Sophie Beck,$^3$ Olivier Gingras,$^3$ Antoine Georges,$^{4,3,5,6}$  and Jernej Mravlje$^{1,7}$\\[0.5em]

\textit{\small
$^1$Jo\v{z}ef Stefan Institute, Jamova 39, 1000 Ljubljana, Slovenia\\
$^2$Instituto de F\'{\i}sica Rosario (CONICET) and Facultad de Ciencias Exactas, Ingenier\'{i}a y Agrimensura, 
Universidad Nacional de Rosario, 2000 Rosario, Argentina\\
$^3$Center for Computational Quantum Physics, Flatiron Institute, New York 10010, USA\\
$^4$Coll{\`e}ge de France, 11 place Marcelin Berthelot, 75005 Paris, France\\
$^5$CPHT, CNRS, {\'E}cole Polytechnique, Institut Polytechnique de Paris, Route de Saclay, 91128 
Palaiseau, France\\
$^6$Departement of Quantum Matter Physics, University of Geneva, 24 quai 
 Ernest-Ansermet, 1211 Geneva, Switzerland\\
$^7$Faculty of Mathematics and Physics, University of Ljubljana, Jadranska 19, 1000 Ljubljana, Slovenia
}

\vspace{2em}
\end{center}

\twocolumngrid
\setcounter{figure}{0}
\renewcommand\thesection{S\arabic{section}}
\renewcommand{\thefigure}{SM\arabic{figure}}

\section{Generalized Drude description}
\subsection{Raman susceptibility}
Raman susceptibility can be discussed in terms of the generalized Drude model, analogously to the more familiar optical conductivity.
One can define the Raman conductivity  $\sigma(\Omega)= \sigma'(\Omega) + i \sigma''(\Omega)$ with $\sigma'$, $\sigma''$ denoting real and imaginary parts. The Raman conductivity is related to the Raman susceptibility by the relation
$\sigma'(\Omega) = \chi''(\Omega)/\Omega$ \cite{SMShastry1990}. 
As for the optical conductivity, one can express the Raman conductivity in a form similar to the Drude expression (``generalized Drude model'', see e.g. Ref.~\cite{SMBerthod2013})
\begin{equation}\label{eq:gen_Drude}
                \sigma(\Omega)=\chi_0 \frac{\Zopt(\Omega)}{-i\Omega+ \Zopt(\Omega)\Gamma^*(\Omega)},%  \approx \frac{\chi_0^*}{-i\Omega+ 1/\tau^*(\Omega)},
	\end{equation}
where \begin{equation}
\chi_0= \frac{2}{\pi}\int\limits_0^{\Omega_c} \sigma_1(\Omega)d\Omega,
\label{eq:massenh_02}
\end{equation}
The cutoff $\Omega_c$ can be taken to be infinite, but later we will use a finite value. In this expression 
the renormalization factor: 
\begin{equation}
\Zopt(\Omega) = \frac{1}{1 + \lambda(\Omega)} = \frac{m}{m^*(\omega)}
\end{equation}
is the inverse of the Raman mass enhancement and $\Gamma^*(\Omega)$ the Raman scattering rate. 
At low energies, $\Zopt$ tends to a constant and one can write
\begin{equation}\label{eq:gen_Drude2}
                \sigma(\Omega)\approx \frac{\chi_0^*}{-i\Omega+ \Gamma^*(\Omega)},
	\end{equation}
where $\chi_0^* $ and $\Gamma^*$ denote the renormalized Raman `plasma frequency' and the renormalized scattering rate.

\begin{figure}[h]
\begin{center}
\includegraphics*[width=0.95\columnwidth]{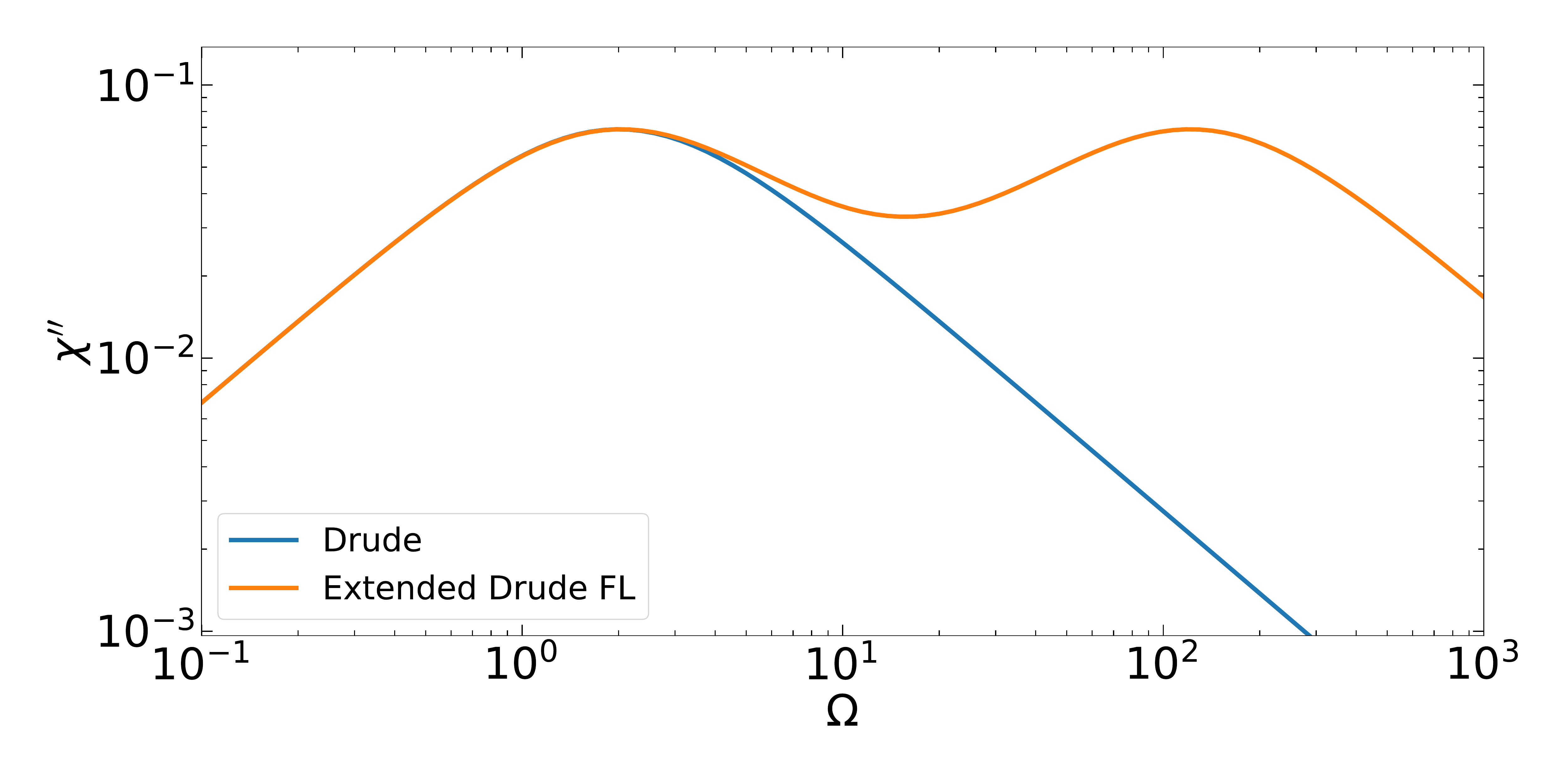}
\caption{
  {Raman susceptibility in the extended Drude model with Fermi liquid $\Gamma^* \propto (\omega^2 +(2 \pi k_B T)^2)$ and low energy Drude behavior $\Gamma^*=$const (blue). } 
}
\label{fig:examples}
\end{center}
\end{figure} 

Hence, 
the imaginary part of Raman susceptibility can be described with   
\begin{equation}
  \chi''_{EDM}(\Omega)=   \Omega  \chi_0^*\frac{\Gamma^* (\Omega)}{\Omega^2 + \left[\Gamma^*(\Omega)\right]^2}
\end{equation}

The Raman susceptibility corresponding to a Drude behavior $\Gamma^* \rightarrow \mathrm{const}$ and to the Fermi liquid behavior $\Gamma^* \rightarrow \alpha \left[ \Omega^2 + (2 \pi k_B T)^2 \right]$ is shown on Fig.~\ref{fig:examples}. Notice the occurrence of the intermediate Fermi liquid minimum in the Fermi liquid case~\cite{SMFreericks2001,SMPhilippe2021}, and the associated two-hump shape, apparent on the log-log plot.

\subsection{Low energy behavior of Raman response in \sro ; the effective masses and scattering rates}

In Ref.~\cite{SMPhilippe2021}, such a generalized Drude analysis was applied to the experimental Raman spectra and 
the reported effective mass enhancement was found to be larger in the \BTg channel as compared to the \BOg one. 
We show here that the same conclusion applies to our theoretical results and explain why this is nonetheless 
consistent with our finding that the \BOg channel is dominated by the $xy$ orbital/$\gamma$-sheet which is known to have the 
largest quasiparticle mass.

\begin{figure}[ht]
\begin{center}
\includegraphics*[width=0.9\columnwidth]{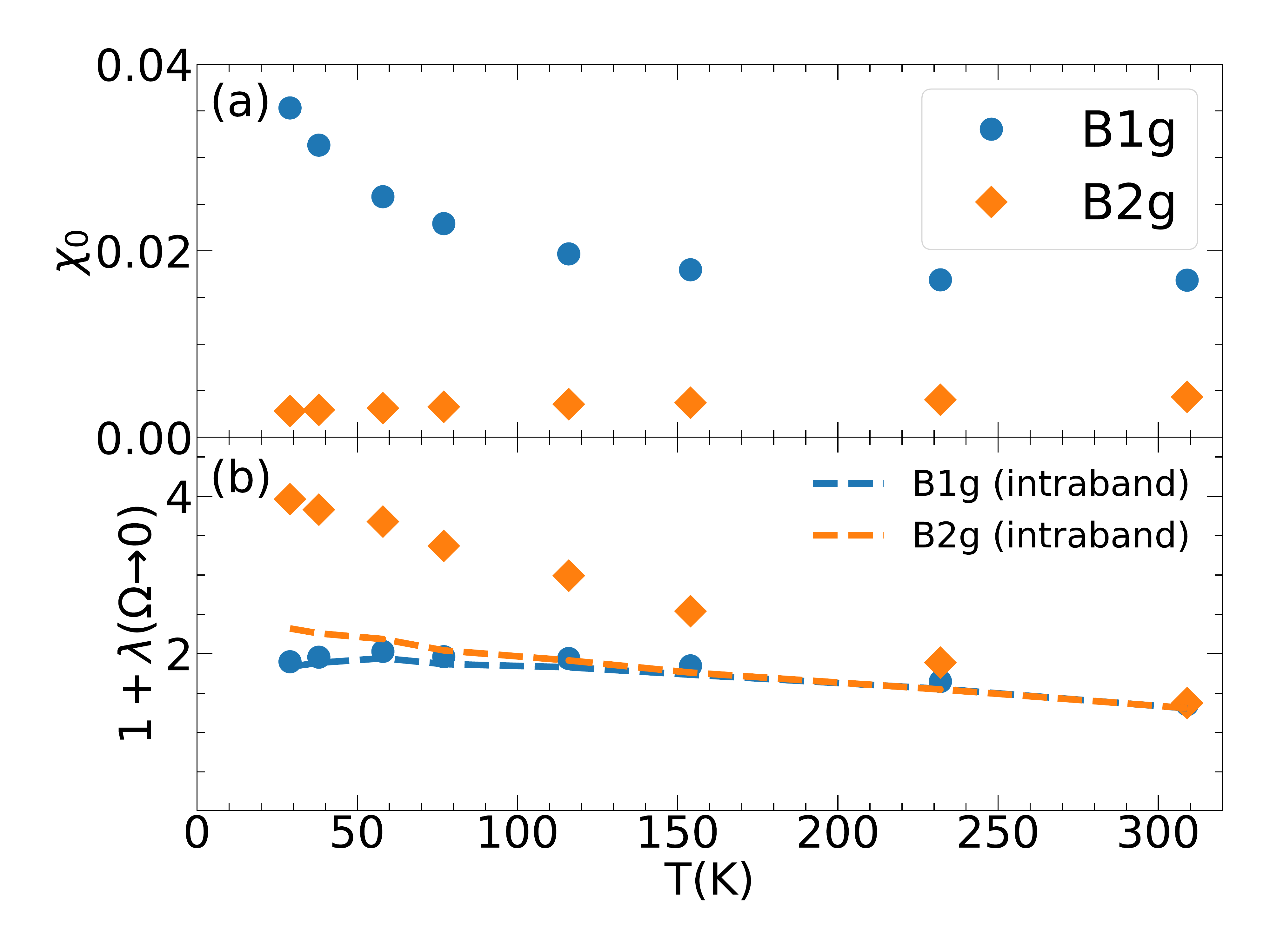}
\caption{(a) Temperature dependence of $\chi_0$ in atomic units. (b) The mass-enhancement factor ($1+\lambda$). In dotted line the mass-enhancement factor if only intraband contributions are considered. We used $\Omega_c=500\,$meV. }
\label{fig:lowfreq}
\end{center}
\end{figure}

The mass-enhancement factor corresponding to the Raman response can be calculated from
\begin{equation}
1+\lambda(\Omega)=\chi_0 \frac{\chi'(\Omega)}{|\chi(\Omega)|^2}.
\label{eq:massenh_2}
\end{equation}
Fig.~\ref{fig:lowfreq} (a) presents the behavior of $\chi_0$ as a function of $T$. Similarly to what is found in measurements of Ref.~\cite{SMPhilippe2021} $\chi_0$ in the \BOg drops strongly with temperature, whereas the \BTg depends on $T$ less (our results show a weak increase with $T$). Fig.~\ref{fig:lowfreq} (b) shows the behavior of the mass enhancements. Weaker mass enhancement is found in the \BOg channel. Additionally, we show the result of a calculation in which we neglect the interband terms. This demonstrates that the stronger mass enhancements in \BTg are caused by the interband terms. 
In particular, the large interband contribution leads to an underestimation of relative weight of 
the low-energy response when applying a generalized Drude analysis, hence leading to an overestimation of the 
effective mass enhancement.

\begin{figure}[ht]
\begin{center}
\includegraphics*[width=0.9\columnwidth]{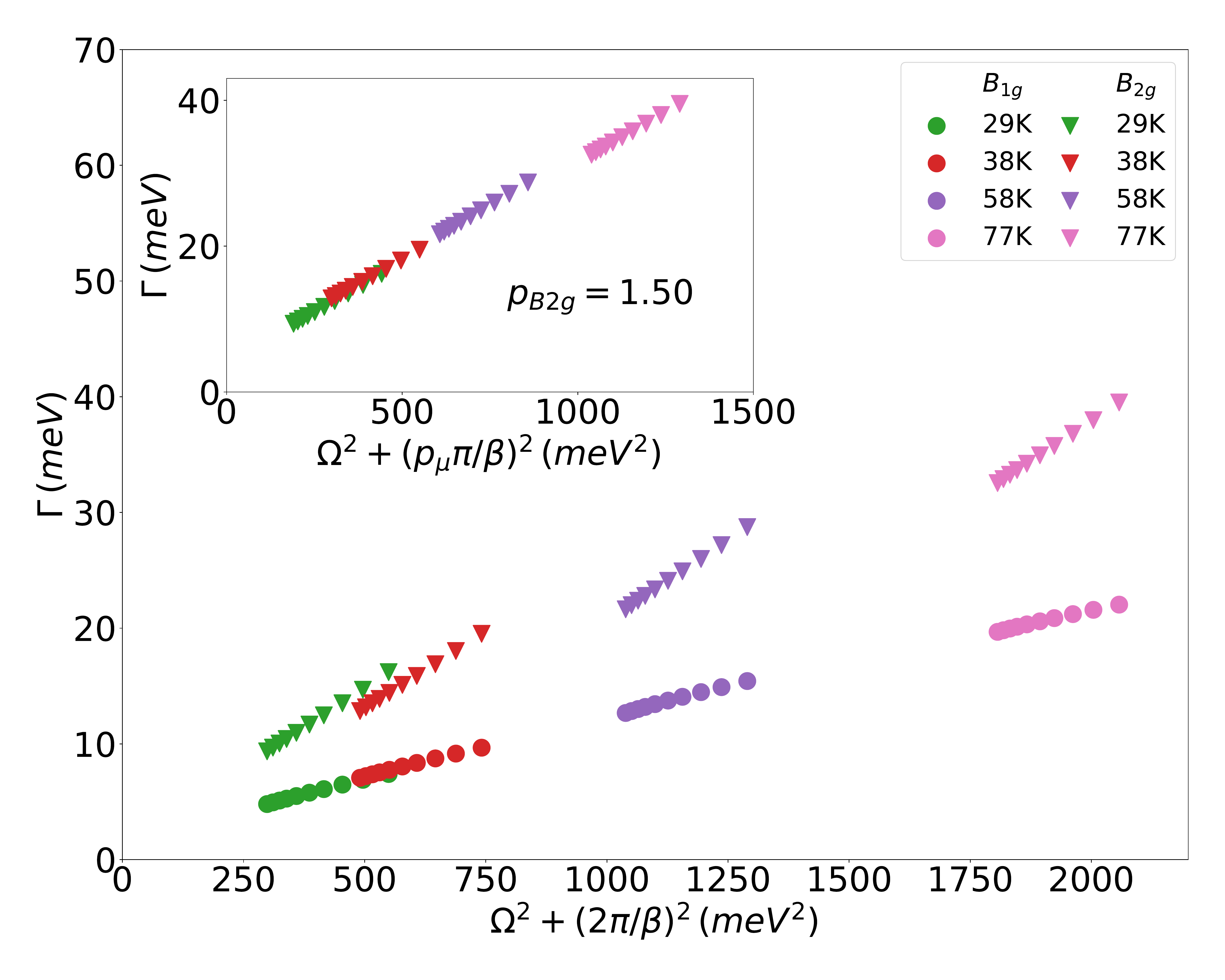}
\caption{Raman relaxation rate in \BOg and \BTg channels (circles and triangle, respectively) vs. $(\Omega^2 + (p \pi k_B T)^2$. In main panel we use $p=2$ and canonic Fermi liquid behavior is indicated by linear behavior.  In \BTg the scattering rate is not linear in  $[\Omega^2 + (p \pi k_B T)^2]$ for $p=2$. Inset demonstrates the data collapse to a linear function when $p=1.5$ is used, instead. }
\label{fig:Gamma_scaling}
\end{center}
\end{figure}

On Fig.~\ref{fig:Gamma_scaling} we plot the Raman scattering rate
\begin{equation}
\Gamma(\Omega)=\chi^0 \Omega \frac{ \chi''(\Omega)}{|\chi(\Omega)|^2},\label{eq:relaxrate_2}
\end{equation}
vs. $\Omega^2 + (p \pi k_B T)^2$, to demonstrate the Fermi liquid behavior at low temperatures and energies. One sees the data obeys the Fermi liquid expectations with the Ghurzi factor $p=2$ for the \BOg channel, but a smaller value $p=1.5$ must be used for the \BTg channel.

\section{Self-energies}
\begin{figure}[h]
\begin{center}
\includegraphics*[width=0.9\columnwidth]{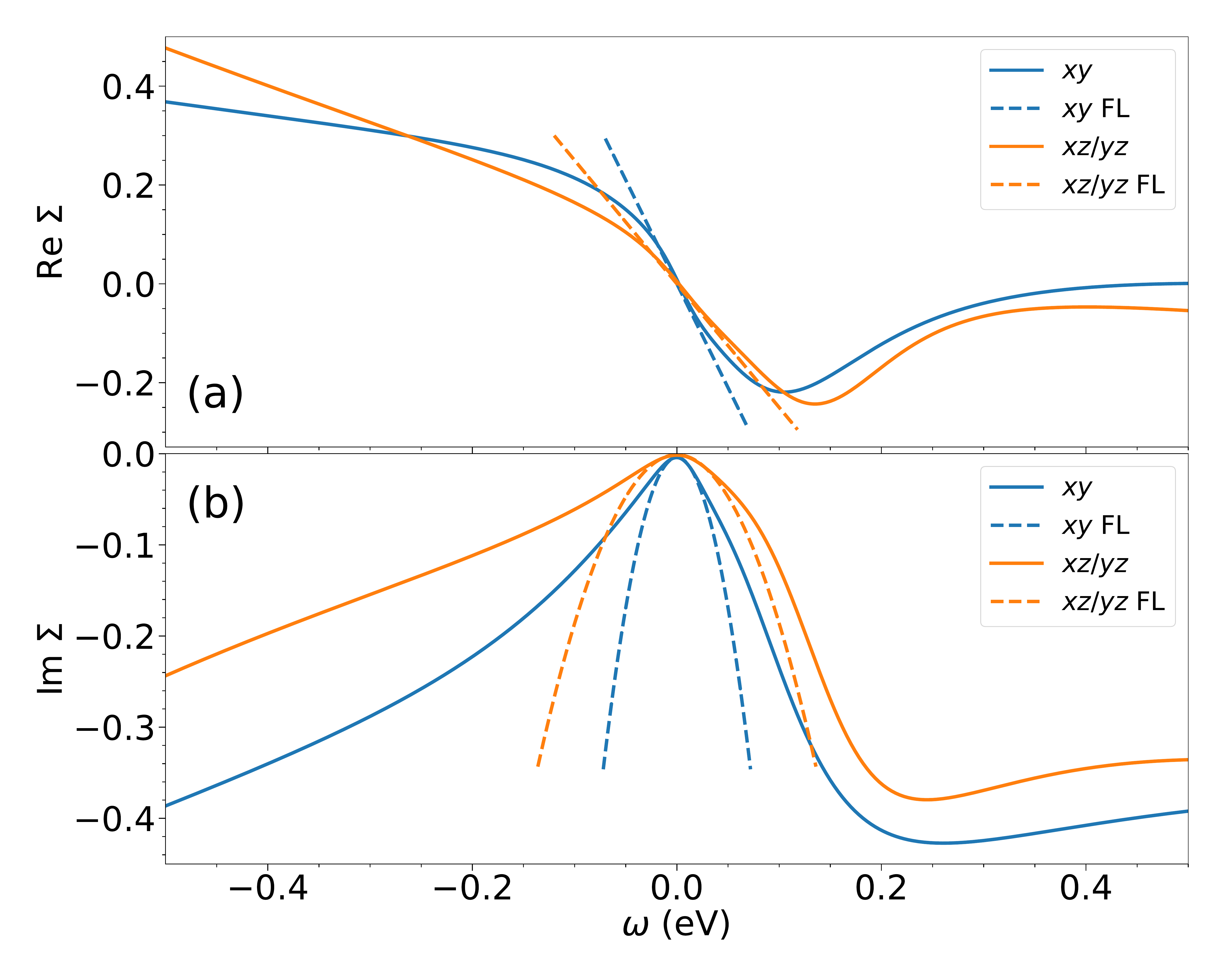}
\caption{
  Real (a) and imaginary (b) part of the self-energy for the orbital $xy$ (blue) and $xz/yz$ (orange) at $29\,$K. Fermi liquid low energy fits are indicated (dashed).  
}
\label{fig:selfenergy}
\end{center}
\end{figure}

The self-energies used in the calculations for temperature $T=29\,$K are shown on Fig.~\ref{fig:selfenergy}. The Fermi liquid  (FL) linear-in-$\omega$ dependence in Re$\Sigma$ and quadratic dependence in Im$\Sigma$ $\propto (\omega^2 + \pi^2 k_B^2 T^2) $, are also indicated.
Notice abrupt change of slope in Re$\Sigma$ at $\omega=0.1\,$eV. This abrupt `unrenormalization' linked to Hund metal nature (Sec.~\ref{sec:Hund}) leads to an enhancement of Raman response above this frequency.  

\section{Individual orbital response: Allen formula}
It is of interest to know the Raman response corresponding to individual orbitals. In an 
idealized case where there is no orbital-mixing and the band structure effects are weak (i.e,  the energy dependence of the band transport function $\Phi(\epsilon)= \sum_\mathbf{k} \gamma_\mathbf{k}^2 \delta(\epsilon-\epsilon_k)$  is weak ~\cite{SMBerthod2013}) the Raman response is given by the Allen formula~\cite{SMAllen2015} \begin{equation}
\chi''(\Omega)\propto \mathrm{Im}  \int\limits_{-\infty}^\infty d\omega \frac{f(\omega)-f(\omega+\Omega)}{\Omega+\Sigma^*(\omega)-\Sigma(\omega+\Omega)}.
\label{eq:Allen}
\end{equation}
The Allen formula is expressed in terms of one component of self-energy alone and corresponds to an ``ideal Raman filter''.  

\begin{figure}[h]
\begin{center}
\includegraphics*[width=0.9\columnwidth]{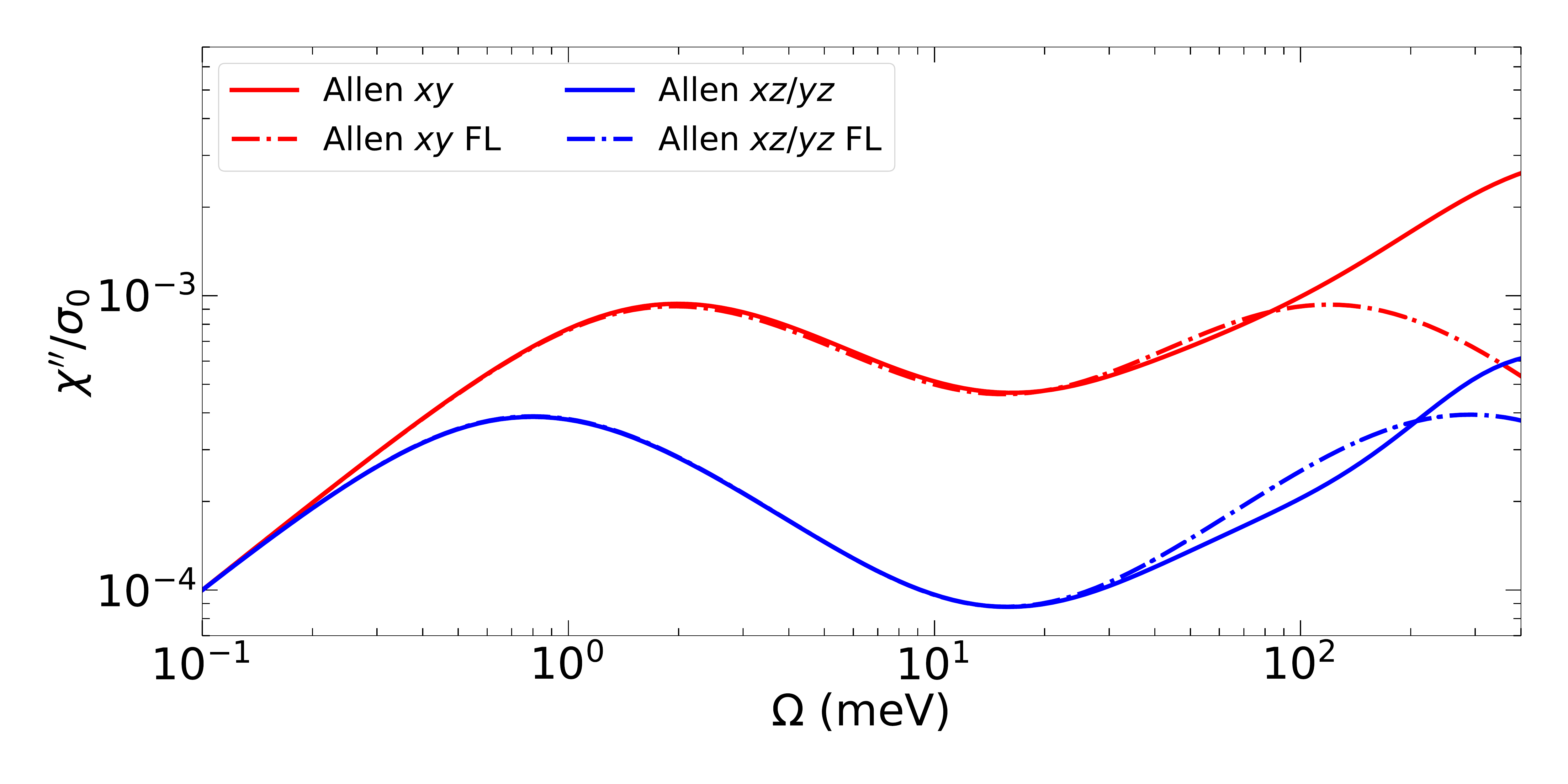}
\caption{
  Normalized Raman response using Allen formula Eq. \ref{eq:Allen}) for $xy$ (red) and $xz/yz$ orbital (blue). The results using corresponding Fermi liquid fits are also shown (dashed-dotted line).
}
\label{fig:allenSM}
\end{center}
\end{figure}

 In Fig.~\ref{fig:allenSM}, we show the Allen formula results (normalized by $\sigma(0) =\chi''(\Omega)/\Omega)|_{\Omega\rightarrow 0}$  so that the data match for small $\Omega$) calculated with the DMFT self-energies for the two obitals (full) and compare the results also to what one obtains using a Fermi liquid fit to the two self-energies (dashed-dotted). In $xz$ orbital one sees a narrower Drude-peak followed by a more pronounced Fermi liquid intermediate minimum than what one sees for $xy$. This is consistent with a more coherent response in the $xz$ orbital known from earlier work~\cite{SMMravlje2011,SMStricker2014}. Except for the small deviations at $30\,$meV, FL describes the behavior pretty well all the way up to $100\,$meV for the $xy$ and $200\,$meV for the $xz$ orbital.  Because these results correspond to an `ideal Raman filter' that would selectively probe each of the two orbitals, and because the 80meV hump feature discussed in the main text is not observed in either of the two, its occurrence cannot be explained in terms of the orbital-physics/correlations. At higher frequencies the actual response strongly exceeds the Fermi liquid response, which is a signature of Hund metals as discussed next.

\begin{figure}[ht]
\begin{center}
\includegraphics*[width=0.9\columnwidth]{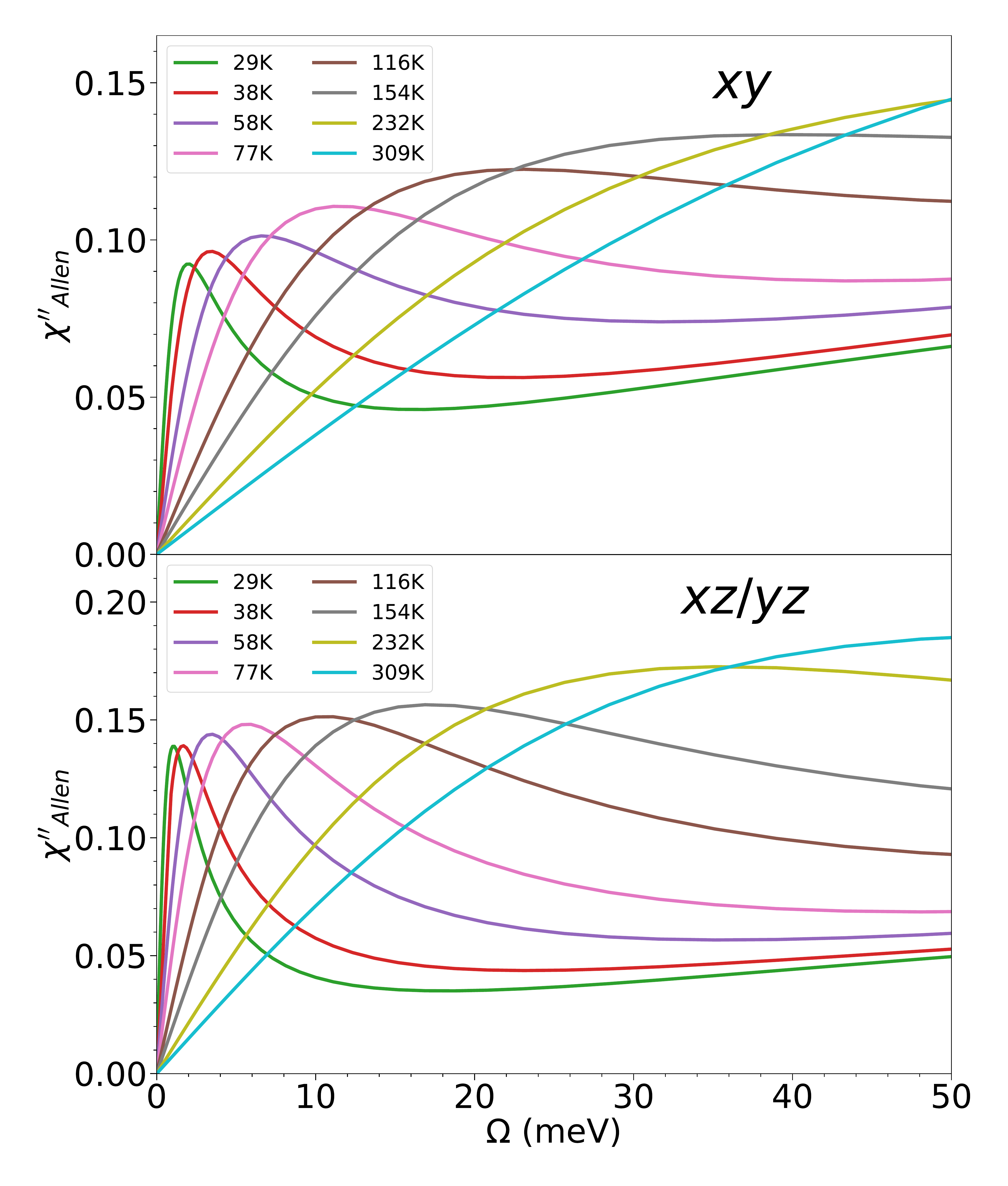}
\caption{Allen formula response for orbital $xy$ (top) and $xz/yz$ (bottom) for different temperatures. The Fermi liquid self-energy is $\Sigma_{FL}(\omega)=(1-1/Z_m)\omega -i A_m (\omega^2 + \pi^2 k_B^2 T^2)$ where, $Z_{xy}=0.19$ and $A_{xy}=66$ , and $Z_{xz/yz}=0.29,\,A_{xz/yz}=18.5$. Same values were used in Fig.~\ref{fig:FL_NFL}
of the main text.}
\label{fig:allen_v_T}
\end{center}
\end{figure}

In Fig.~\ref{fig:allen_v_T} we present the Allen formula results for several temperatures for $xy$ (top) and $xz$ (bottom) orbitals. With the exception of a broader Drude peak found for the $xy$ case, the two sets of data behave similarly.  

\section{Hund metal signature}
\label{sec:Hund}
In this section, we relate the enhancement seen in the Raman response at frequencies above $0.1\,$eV to the shape of single particle density of states that we argue is characteristic of Hund metals. To demonstrate this we also performed calculations using Sr$_2$RuO$_4$ band structure but at increased occupancy of $N=5$ electrons in t$_{2g}$ orbitals, which provides a reference where no Hund metal effects are present. In those calculations we increased the Hubbard interaction parameter to $U=4.2\,$eV so that the resulting mass enhancement is the same as that of the $N=4$ calculation. 

\begin{figure}[ht]
\begin{center}
\includegraphics*[width=0.98\columnwidth]{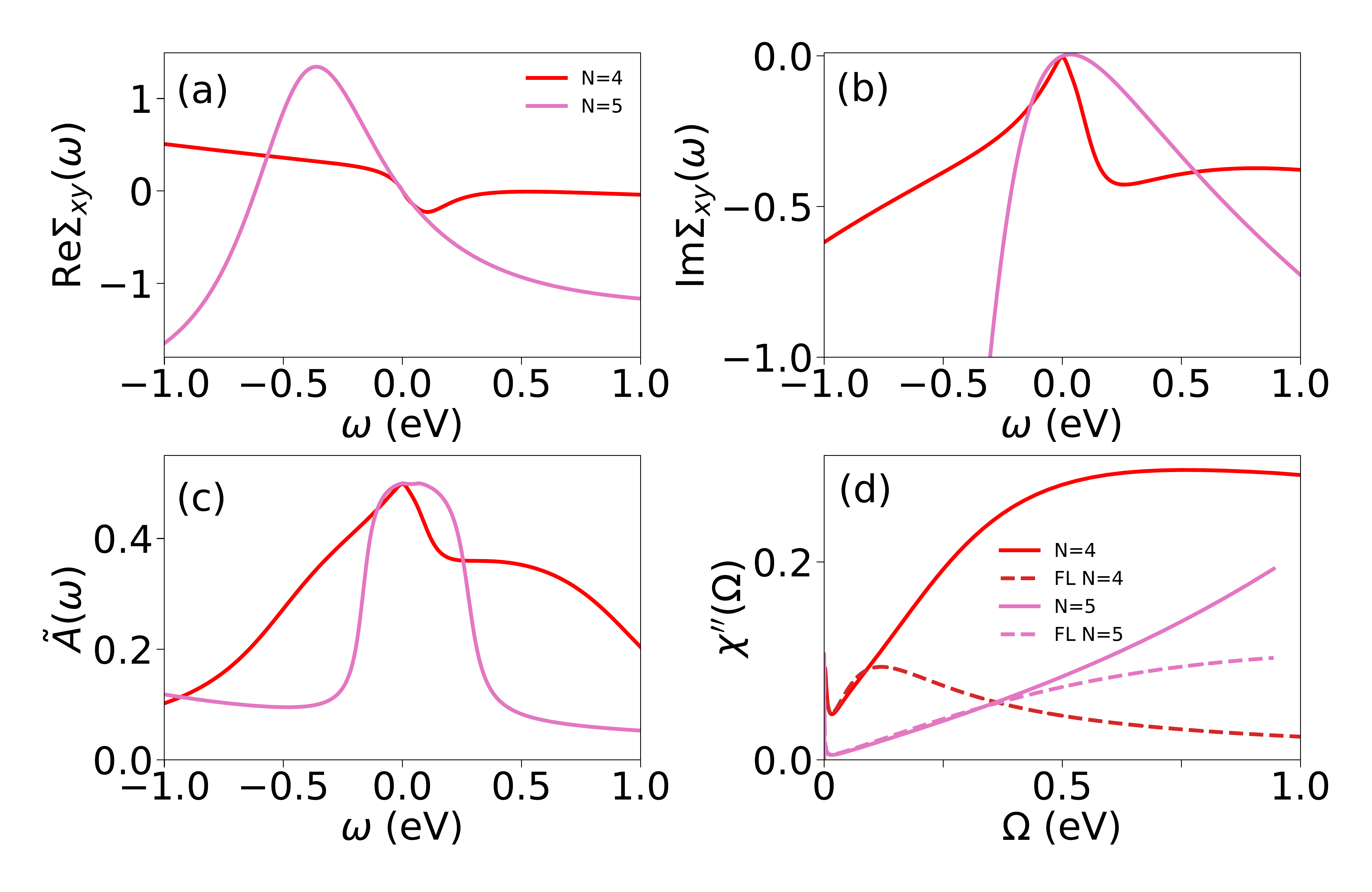}
\caption{ Calculation with the self-energies obtained from DMFT for $xy$ orbital and $\beta=400$ for \sro in red ($N=4$), and using the same band-structure but considering increased occupancy $N=5$ in pink. The real and imaginary part of the self-energy are compared in (a) and (b) respectively.  In (c) the corresponding proxy-spectral functions $\tilde{A}$ are compared. In (d) the corresponding Raman response $\chi''(\Omega)$ obtained from Allen formula (full) are shown and the results are additionally compared to the response of the corresponding Fermi liquid (dashed lines).}
\label{fig:sidepeak}
\end{center}
\end{figure}

On Fig.~\ref{fig:sidepeak}~(a) we show $\mathrm{Re} \Sigma(\omega)$ for $xy$ orbital at $29\,$K. In Hund metal case (red) the strong deviation from low-energy linear behavior in Re$\Sigma$ occurs at a low energy $0.1\,$eV. Conversely in the $N=5$ result (pink)  deviations occur at larger energies only. 

To see how these differences affect the spectral function it is convenient to consider  a proxy quantity $\tilde{A}(\omega)$  
\begin{equation}
\tilde{A}(\omega) = -(1/\pi) \mathrm{Im}  \int d\varepsilon\, \tilde{\rho}(\varepsilon)[ \omega - \varepsilon - \Sigma(\omega) ]^{-1}
\label{eq:Atilde}
\end{equation}
that gives the density of states of an auxiliary reference single-orbital problem. For this auxiliary problem we take a flat noninteracting DOS with $\tilde{\rho}(\varepsilon)$ being a nonzero constant in the range $[-1\,\mathrm{eV},1\,\mathrm{eV}]$ and vanishing elsewhere. Using such featureless DOS is convenient in order to assure that all features in $\tilde{A}$ are caused by correlation effects. 
The results are shown on Fig.~\ref{fig:sidepeak} (c). The $N=5$ calculation has a shape that is characteristic of standard correlated metals with a narrow quasiparticle peak with a shape resembling narrowed bare DOS  whereas $N=4$ calculation displays a richer structure characteristic of Hund metals with slow decrease of the spectral function at negative energies and a side-hump feature at positive energies. 

 We use the Allen formula (Eq. \ref{eq:Allen}) and calculate also the corresponding response $\chi''$. In Fig. \ref{fig:sidepeak} (d) we compare the results also against the Fermi liquid fits [distinct for the $N=4$ and $N=5$ calculation because the curvatures of the fitted imaginary parts of self-energies are distinct as seen in Fig.~\ref{fig:sidepeak} (b)]. The Fermi liquid behavior is followed to  much larger frequencies for the $N=5$ calculation and the deviations from the Fermi liquid are much smaller than that seen in the $N=4$ calculation, which demonstrates that strong increase of Raman response in the $0.1-0.5\,$eV range discussed in the main text is a signature of the Hund metal.

\section{Raman response for trial self-energies}
\begin{figure}[ht]
\begin{center}
\includegraphics*[width=0.9\columnwidth]{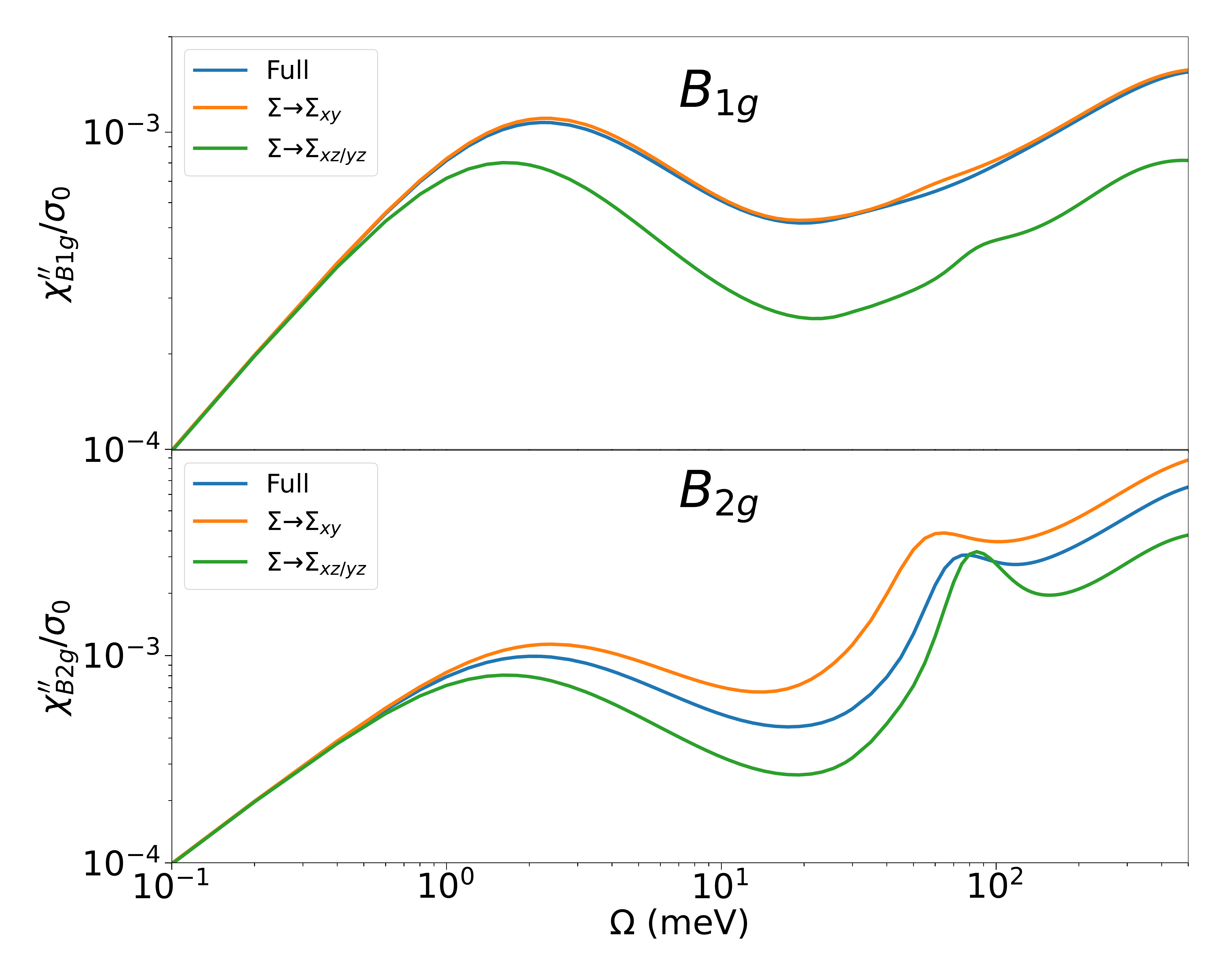}
\caption{Raman response for \BOg (top) and \BTg (bottom) channels for $29\,$K. The calculation using the self-energy of the Sr$_2$RuO$_4$ (blue line) is compared to the case where the self-energy for all orbitals is given by that corresponding to the orbital $xy$ (orange) or the orbital $xz/yz$ (green).}
\label{fig:test_selfenergies}
\end{center}
\end{figure}

To disentangle different physical effects that contribute to the calculated Raman response it is convenient to perform calculations evaluating the Raman response with Eq.~\ref{eq:transp_dist}
 from the main text, but substituting the self-energies with different trial values $\Sigma \to \Sigma_{trial}$. 
In Fig.~\ref{fig:test_selfenergies} we compare the full Raman response  (blue line) with that of a trial calculation where all orbital components of the self-energy were set to that of the $xy$-orbital  (orange line) and to the $xz/yz$ orbital (green line). 
 For the \BOg (top panel) we observe matching between orange and blue lines, indicating that \BOg is mainly contributed by the states involving  the $xy$ orbital. Conversely, in \BTg the full result is in the middle of the two-trial calculations, suggesting both $xz/yz$ and $xy$ orbitals contribute similarly to the response.

\section{Interband contributions}
In Eq.~\ref{eq:transp_dist}
 in the main text, the response has intra-band, inter-band, and mixed contributions. Namely, both spectral functions and Raman vertices have interband components. The spectral density in the band basis is ${\cal A}_k=\left( \omega + \mu - \epsilon_k + \Sigma^\mathrm{band} \right)^{-1}$ where $\Sigma^\mathrm{band} = P \Sigma^\mathrm{orb}(\omega) P^\dagger$. $\Sigma^\mathrm{band}$ and $\Sigma^\mathrm{orb}$ are the self-energy in the band basis and orbital basis respectively and $P$ is the transformation matrix between the two basis. 
It is convenient to ne  ${\cal A}_k$ is diagonal in band basis (this is strictly correct in the case of $\Sigma^\mathrm{orb}\propto \mathbb{I}$). Under this assumption, the Raman response separates into terms that come from the diagonal terms of the vertex $\gamma^\mu_\mathbf{k}$ (the intraband terms) and from the off-diagonal terms (the interband contributions). Namely,

\begin{equation}
\begin{split}
\mathrm{Tr}  \, \gamma^\mu_\mathbf{k} \, {\cal A}_\mathbf{k}(\omega) \, \gamma^\mu_\mathbf{k} \, {\cal A}_\mathbf{k}(\omega + \Omega) \to \qquad \qquad \qquad \qquad \\
~\begin{cases}
\mathrm{intraband:} & \sum_\nu \gamma^\mu_{\nu\nu} \, {\cal A}_{\nu\nu}(\omega+\Omega) \, \gamma^\mu_{\nu\nu} \, {\cal A}_{\nu\nu}(\omega) \\
\mathrm{interband:} & \sum_{\nu\neq \tau} \gamma^\mu_{\nu\tau} \, {\cal A}_{\tau\tau}(\omega+\Omega) \, \gamma^\mu_{\tau\nu} \, {\cal A}_{\nu\nu}(\omega) \\
\end{cases}
\end{split}
\label{eq:intrainter}
\end{equation}

\begin{figure}[b]
\begin{center}
\includegraphics*[width=0.9\columnwidth]{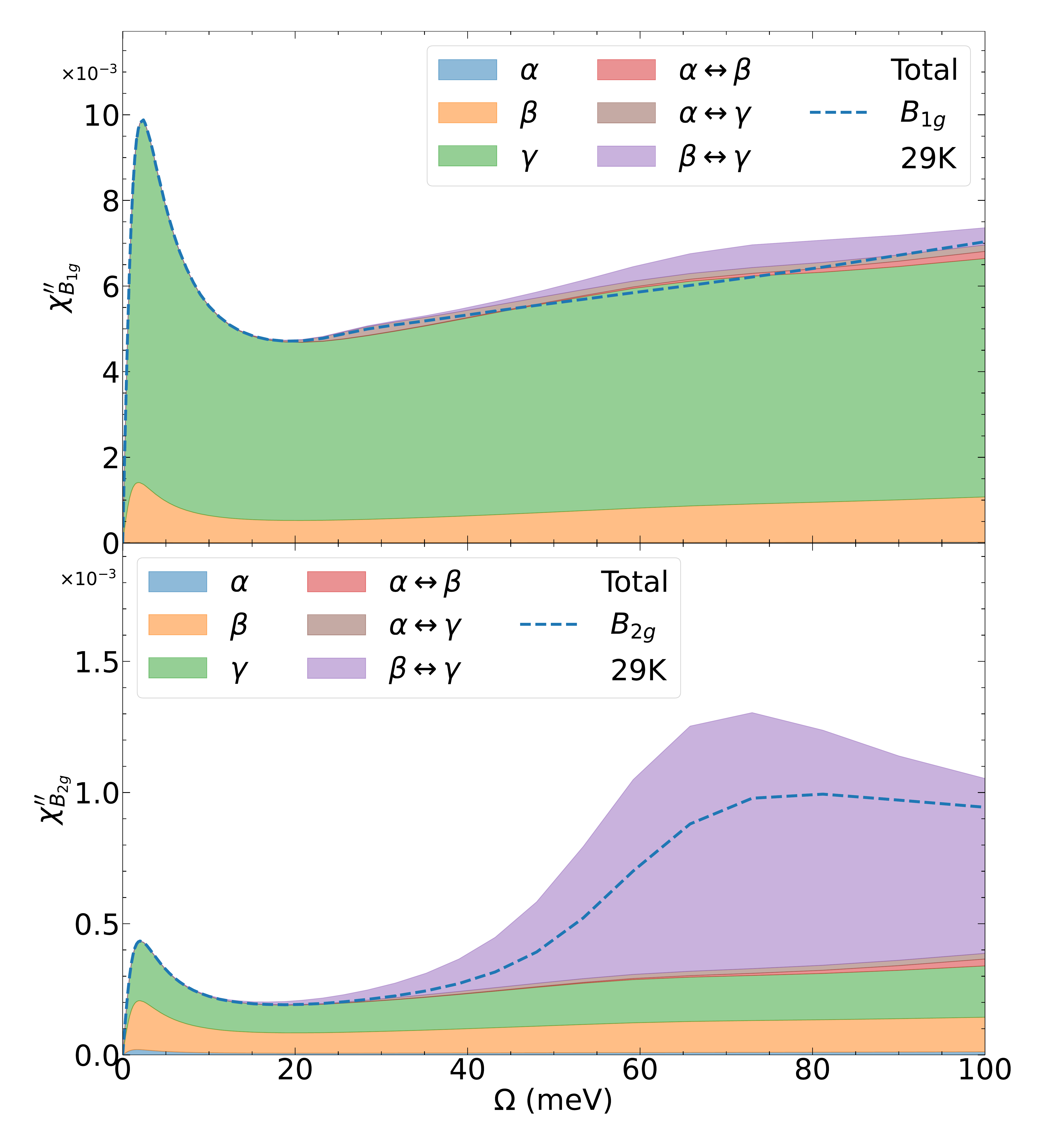}
\caption{Contributions to the Raman response for the \BOg channel (top) and \BTg channel (bottom) at 29K presented in terms of additive contributions (shading) in atomic units. }
\label{fig:interband_B2g}
\end{center}
\end{figure}

Such decomposition is used to analyze the intraband and interband contributions to Raman susceptibility presented in the form of stacked area chart on Fig.~\ref{fig:interband_B2g}. 
Different intraband contributions (blue, orange, and green shading for respectively $\alpha, \beta, \gamma$ bands) and interband contributions (red, brown, violet) are indicated. One sees that for \BOg channel, the majority of the response comes from intraband terms and in particular the $\gamma$ band. Conversely, for \BTg above $30\,$meV  the interband $\beta \leftrightarrow \gamma $ transitions start dominating. Notice that the interband transitions are maximal close to $75\,$meV in both channels and that their absolute magnitude is actually comparable in the two channels. The qualitative distinction in the behavior between the two channels comes from the fact that the intraband terms are much smaller in \BTg, which in turn is due to the fact that the nearest neighbor hoppings do not contribute in that channel. 
Note also that the sum of the contributions exceeds somewhat  the total Raman response (blue dashed line), which is due to the neglected off-diagonal parts of the spectral functions in the band basis in this analysis.
Namely, if one retains those there are  additional mixed contributions to the response which are of the form 
\begin{equation}
\label{eq:mixed}
~\begin{cases}
\mathrm{mixed\, 1:} & \sum_{\nu \neq \tau} \gamma^\mu_{\nu \nu} \, {\cal A}_{\nu\tau}(\omega+\Omega) \, 
\gamma^\mu_{\tau\tau} \, {\cal A}_{\tau\nu}(\omega) \\
\mathrm{mixed\, 2:} & \sum_{\nu\neq \tau} \gamma^\mu_{\nu\tau} \, {\cal A}_{\tau\tau}(\omega+\Omega) \, \gamma^\mu_{\tau\tau} \, {\cal A}_{\tau\nu}(\omega) \\
\mathrm{mixed\, 3:} & \sum_{\nu \neq \nu' \neq \tau  \neq \tau' } \gamma^\mu_{\nu\nu'} \, {\cal A}_{\nu'\tau}(\omega+\Omega) \, \gamma^\mu_{\tau\tau'} \, {\cal A}_{\tau'\nu}(\omega) \\
\end{cases}
\end{equation}
and give a negative contribution to the response.

\section{Expression for A$_{1g}$+B$_{2g}$ vertex}
A$_{1g}$+B$_{2g}$ is probed by both incident and reflected light being polarized inplane at 45 degrees to the nearest neighbor Ru-O bond, i.e. in $110$ direction (using conventional unit cell notation). The corresponding vertex reads
\begin{equation}
\gamma^{A_{1g}+B_{2g}}_\mathbf{k}=\frac{\partial^2 H^{(W)}}{\partial k_x \partial k_y} +\frac{1}{2}\left( \frac{\partial^2 H^{(W)}}{\partial k_x^2 \hfill} + \frac{\partial^2 H^{(W)}}{\partial k_y^2 \hfill}\right).
\end{equation} 

\section{Vertex corrections}
\begin{figure}[t]
    \centering
    \includegraphics[width=\linewidth]{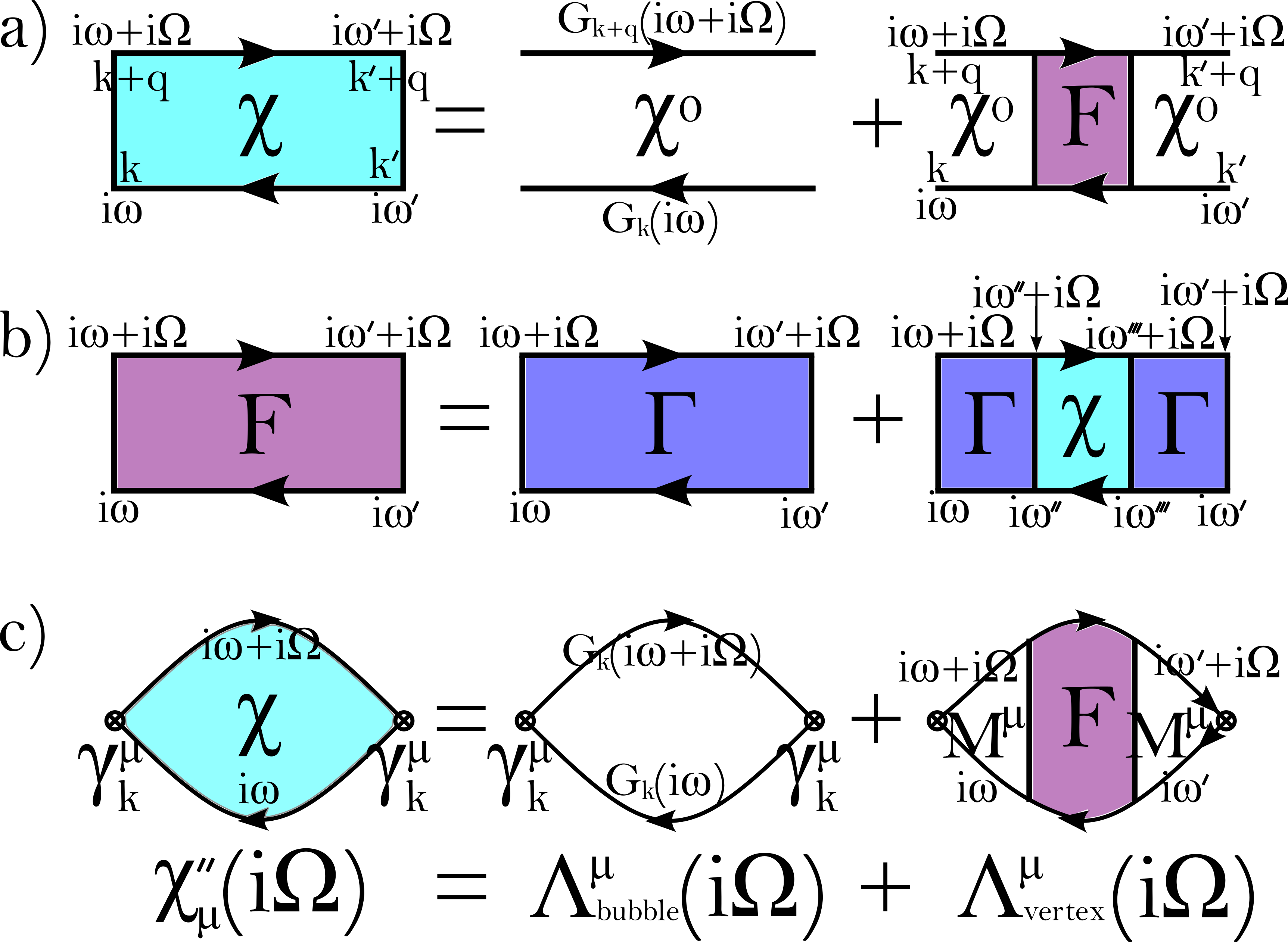}
    \caption{Diagrammatic representation of the many-body vertex corrections to the Raman response. (a) Vertex corrections to the lattice particle-hole susceptibility $\chi$. (b) Reducible vertex $F$ expressed in terms of the irreducible vertex $\Gamma$. (c) Full Raman responses (or optical conductivity, if the Raman vertex is substituted by that of velocity $\gamma \rightarrow v $). We denote $M=G \gamma G$.  Summation over momentum $k$ and fermionic frequencies $\omega$ are implied.}
    \label{fig:raman_vertex}
\end{figure}

It is well known that for inversion-symmetric systems  the vertex corrections (VCs) to optical conductivity in DMFT vanish~\cite{SMKhurana1990,SMGeorges1996}. This is because in DMFT, the many-body vertex (i.e. the object $F$ depicted in diagrams shown on  Fig.~\ref{fig:raman_vertex}, not to be confused with the Raman vertex $\gamma$) is local  and VCs are then given by terms of the form $\sum_\mathbf{k} v_\mathbf{k} G_\mathbf{k} G_\mathbf{k}$ which vanish by symmetry (namely, velocity $v_\mathbf{k}$ is odd under $\mathbf{k}\rightarrow -\mathbf{k}$, whereas the Green's function $G_\mathbf{k}$ is even). The situation in multi-band problems is more subtle, but one can argue that under some limitations the vertex correction vanish there, too~\cite{SMDang2015PRL}. We checked that in Sr$_2$RuO$_4$, the Green's function matrix (in orbital basis) has even parity and velocity matrix is odd under inversion, so the argument applies and the VCs to optical conductivity in DMFT vanish. 

What about Raman response? In cluster extensions of DMFT the  VCs were shown to be relevant~\cite{SMLin2010,SMLin2012,SMGull2013}, but the situation in pure DMFT was to our knowledge not discussed before. 

In the Raman response, the role of velocity is taken by Raman vertex $\gamma_\mathbf{k}$, which is not odd under inversion. However, other symmetries could also cause vanishing of  $\sum_\mathbf{k} v_\mathbf{k} G_\mathbf{k} G_\mathbf{k}$.  Consider  a tight-binding model  in 2D with  the band energy  $\varepsilon(k_x,k_y) = t [\cos (k_x) + \cos (k_y)] + t' \cos(k_x) \cos(k_y) $, with $t,t'$ respectively the nearest neighbor  hopping (next nearest neighbor hopping). For such a band, the Raman vertex in \BOg is $\gamma^{\text{B}_{1g}}= 1/2 (\partial^2 \varepsilon /\partial k_x^2 - \partial^2 \varepsilon /\partial k_y^2 ) = t' [\cos (k_y) - \cos (k_x)]/2$ which is odd under reflections across zone diagonals $k_x =\pm k_y$, whereas the Green's functions are even under those reflections. Hence $\sum_\mathbf{k} \gamma^{\text{B}_{1g}}_\mathbf{k} G_\mathbf{k} G_\mathbf{k}$ vanishes.  A similar argument applies to the \BTg channel, where $\gamma^{\text{B}_{2g}}= \partial^2 \varepsilon /\partial k_x\partial k_y = t' \sin(k_x) \sin(k_y)$ and is odd under reflection across $k_x=0$ and $k_y=0$ planes.

\begin{figure}[h]
\begin{center}

\includegraphics*[width=0.95\columnwidth]{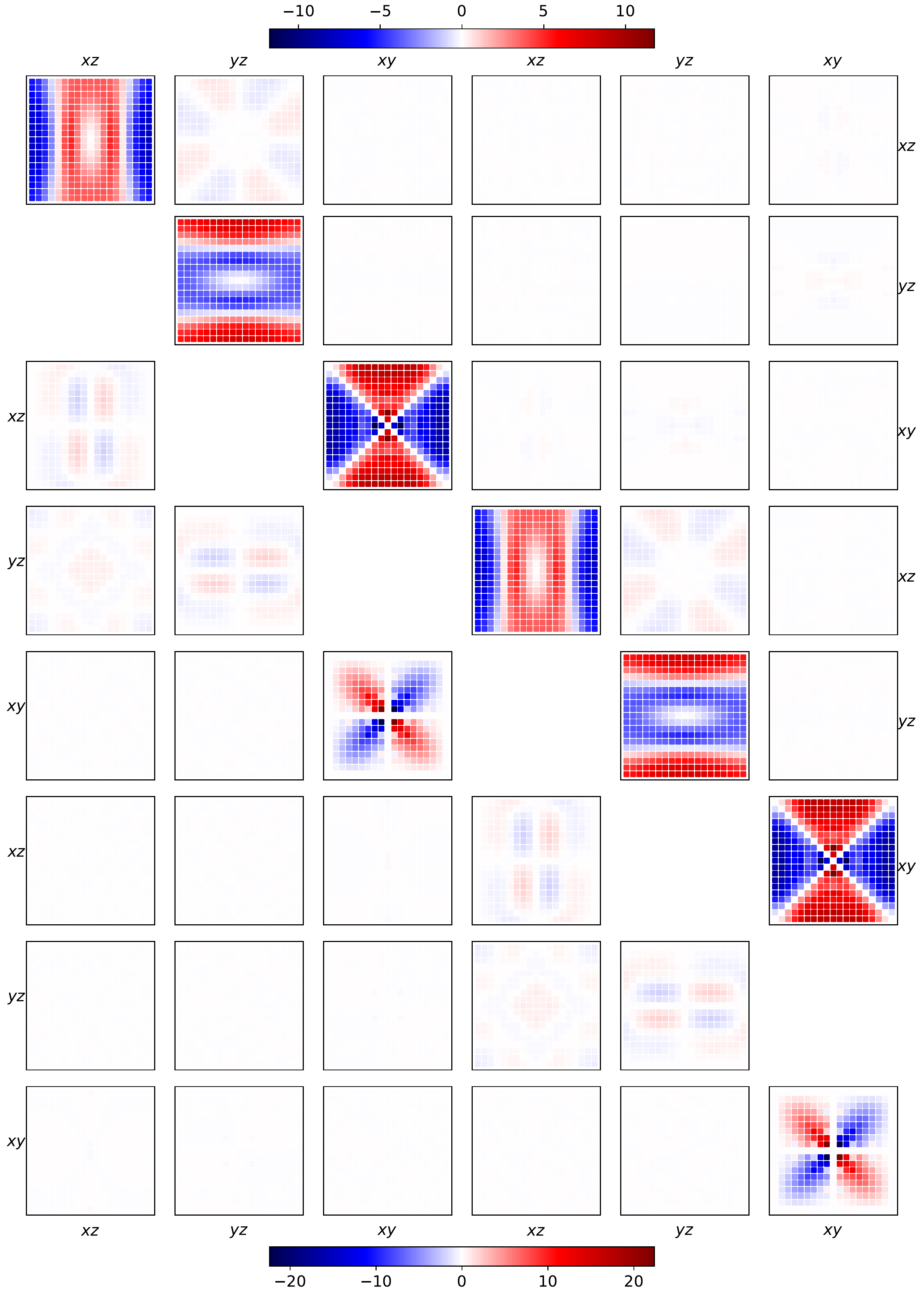}
\caption{Value of the Raman vertex in a plane $k_z=0$. Each orbital index combination $m,m'$ (for the six spin-orbitals) is shown in a separate panel. The scale is kept the same within the same channel for better comparison. The panels in the upper triangle correspond to the \BOg channel, and the one in the lower triangle to the \BTg channel. The values are in atomic units.}
 \label{fig:vertex_B1gB2g}
\end{center}
\end{figure}

\begin{figure}[h]
\begin{center}
\includegraphics*[width=0.95\columnwidth]{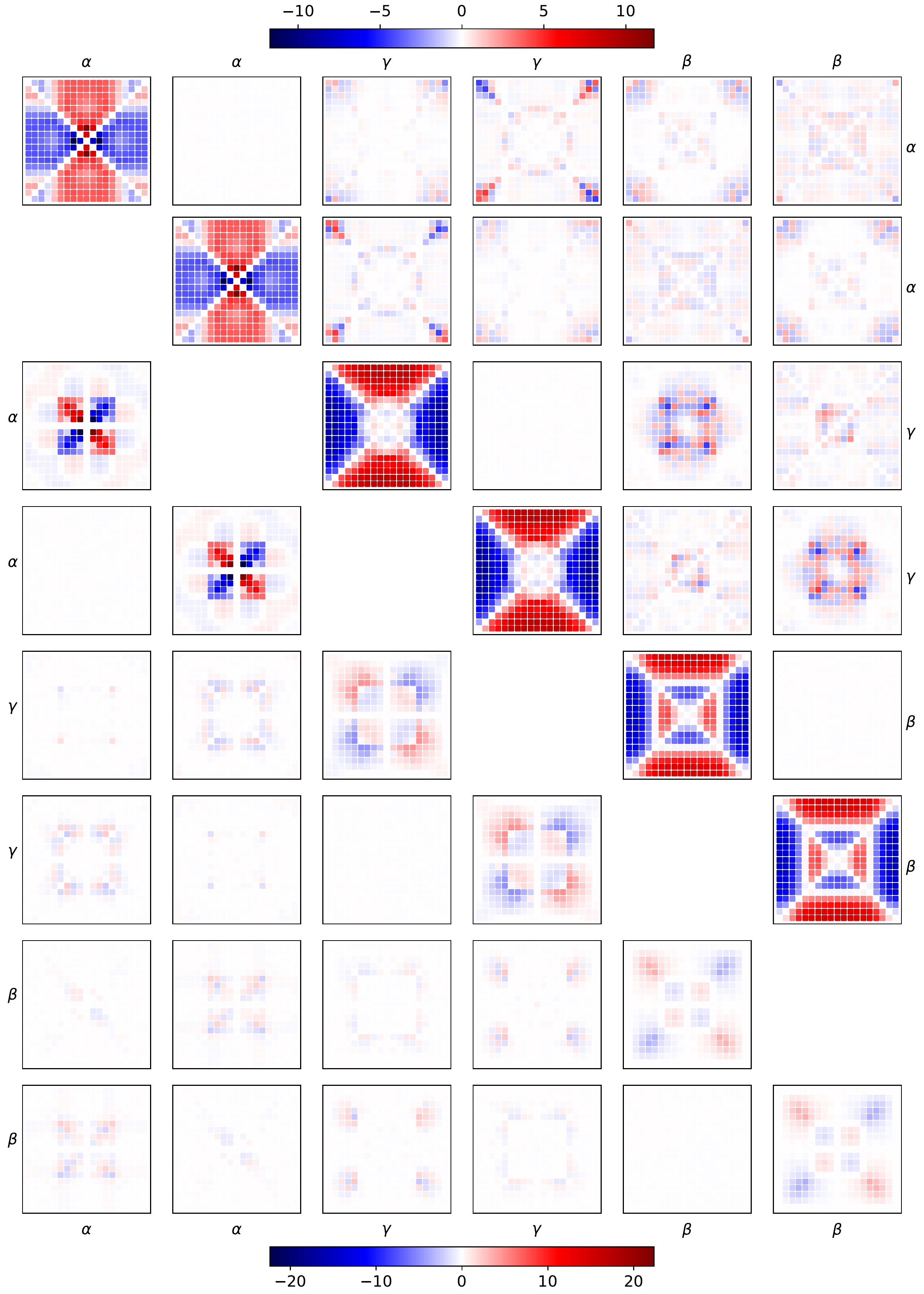}
\caption{Value of the Raman vertex in a plane $k_z=0$. Each band index combination $\nu',\nu'$ (for the six spin-orbital bands) is shown in a separate panel. The scale is kept the same within the same channel for better comparison. The panels in the upper triangle correspond to the \BOg channel, and the one in the lower triangle to the \BTg channel. The values are in atomic units.}
 \label{fig:vertexcorr_B1gB2g_m}
\end{center}
\end{figure}

What about the multi-orbital Sr$_2$RuO$_4$ case? The
momentum dependencies of the Raman vertex for different spin-orbital index combinations are shown on Fig.~\ref{fig:vertex_B1gB2g}. Whereas these matrix elements do reflect symmetries, the momentum sum of $A_k \gamma_k A_k$  (shown on Fig.~\ref{fig:vertexcorr_B1gB2g_m} does not vanish for all spin-orbital channels, so one cannot argue about the vanishing of VCs from those considerations alone.

Therefore, we evaluated the many-body vertex  and calculated the corresponding VCs to the Raman response.
The particle-hole reducible vertex $F$ affects the propagation of particles and holes in the system following the Bethe-Salpeter equation for the particle-hole susceptibility
$\chi = \chi^0 + \chi^0 F \chi^0$ 
where the summations over internal spin, orbital and Matsubara indices are implicit, diagrammatically depicted in Fig.~\ref{fig:raman_vertex}~(a).
The fully reducible many-body vertex can itself be expressed in terms of the irreducible vertex $\Gamma$ as $F = \Gamma + \Gamma \chi \Gamma$, depicted in Fig.~\ref{fig:raman_vertex}~(b).
The full Raman response can then be separated as the sum of the bubble contributions $\Lambda_{\text{bubble}}$ and the VCs $\Lambda_{\text{vertex}}$, depicted in Fig.~\ref{fig:raman_vertex}.

We computed the many-body vertex $\Gamma \rightarrow \Gamma(i\omega_1, i\omega_2, i\Omega)$ from the impurity Bethe-Salpeter equation (where we obtained the full matrix two-particle Green's function using w2Dynamics~\cite{SMkowalski2019}).
$\chi \rightarrow \sum_{k, q}\chi(k, q; i \omega_1, i\omega_2, i \Omega)$ was obtained from the solution of lattice Bethe-Salpeter equation $\chi = \chi^0 + \chi^0\Gamma\chi$ as implemented in TPRF package of TRIQS~\cite{SMstrand2019,SMtprf,SMvanloon2023dual}. We perform the sum over momentum points directly on the susceptibility because the many-body vertex is local in DMFT. Note that we neglected spin-orbit coupling in the construction of these objects.

We evaluated $F$ on Matsubara axis at $\beta=1/(k_B T)=40/$eV, retaining 30 internal Matsubara frequencies and up to $16\times16\times16$ momentum points. We checked that $F$ correctly describes the suppression of the uniform charge susceptibility from the results obtained by considering only the bubble diagram ($2.4$/eV) to that obtained considering the full susceptibility ($1.4/$eV), which we extracted from self-consistent calculations in a uniform field. 

The VCs to the Raman response for the $\mu$ channel are then given by connected part of the Raman correlation function 
\begin{equation}
    \Lambda_\mathrm{vertex}^\mu(i \Omega_n) \equiv M^\mu(i\Omega_n) F(i\Omega_n) M^\mu(i\Omega_n)
\end{equation}
where the sums over the fermionic Matsubara frequencies are implicit and
\begin{equation}
    M^\mu (i \omega_1, i \omega_2, i \Omega) \equiv \beta \sum_k G_k(i \omega_1)  \gamma_k^\mu  G_k(i \omega_2 + i \Omega ) \delta_{\omega_1\omega_2}
\end{equation}
with $\gamma_k^\mu$ the Raman vertex in the $\mu$ channel. We find the connected part to be significantly smaller than the bubble contribution. The ratio $r^\mu=\Lambda^\mu_\mathrm{vertex}(i \Omega_n =0)/\Lambda^\mu_\mathrm{bubble}(i \Omega_n=0)$ is below 1\% for $\mu = $B$_{1g}$, and below 0.01\% for $\mu =$B$_{2g}$.

\begin{figure}[h]
\begin{center}
\includegraphics*[width=0.95\columnwidth]{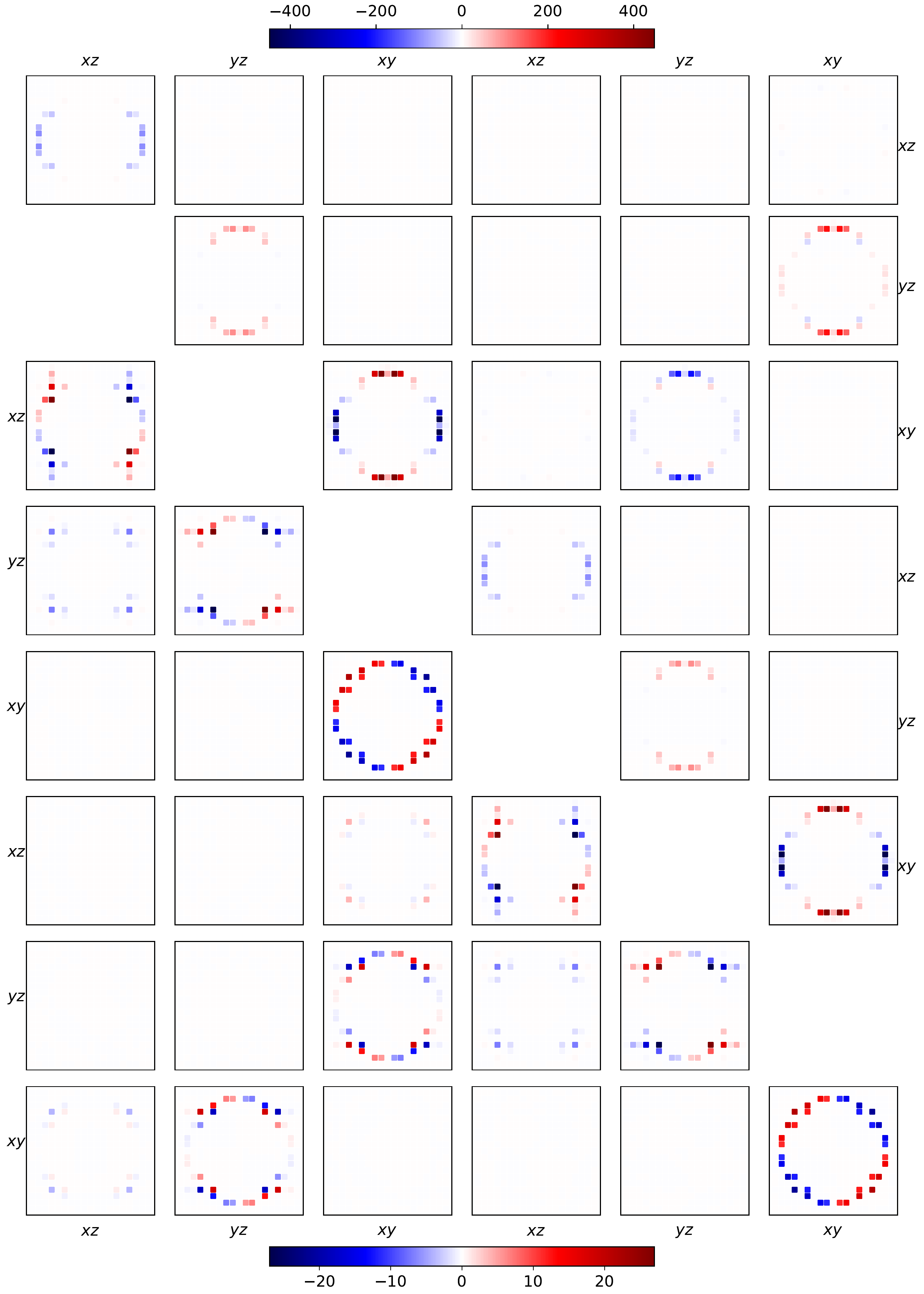}
\caption{Value of $\sum\limits_{m,m'} \gamma_{\mathbf{k}mm'} A_{\mathbf{k} m'm'' }(\omega=0) A_{\mathbf{k} m'''m}(\omega=0)$ in a plane $k_z=0$. Each orbital index combination $m'',m'''$ (for the six spin-orbitals) is shown in a separate panel. The scale is kept the same within the same channel for better comparison.  The panels in the upper triangle correspond to the \BOg channel, and the one in the lower triangle to the \BTg channel. The values are in atomic units.}
\label{fig:vertexcorr_B1gB2g}
\end{center}
\end{figure}

\begin{figure}[h]
\begin{center}
\includegraphics*[width=0.95\columnwidth]{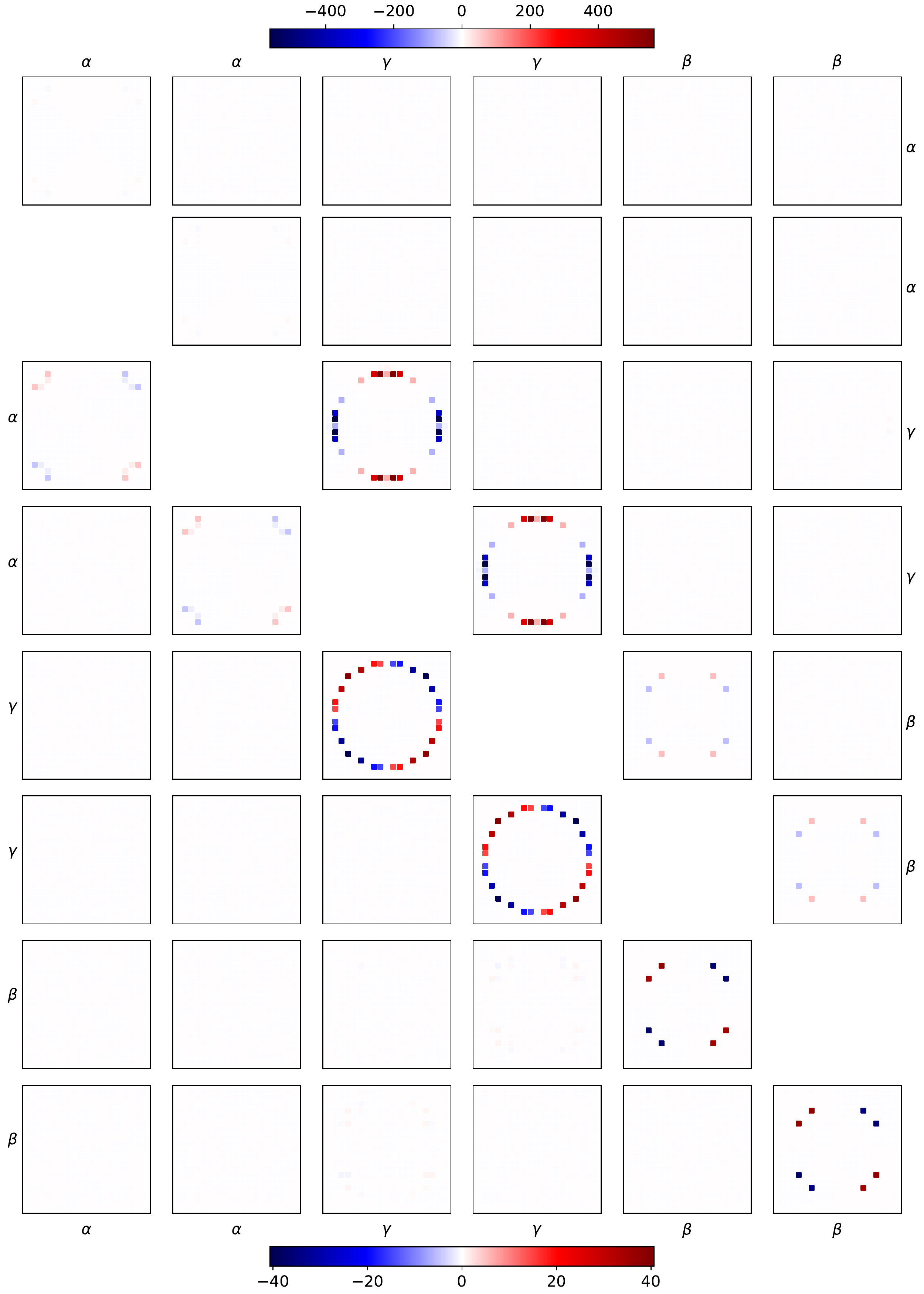}
\caption{Value of $\sum\limits_{\nu,\nu'} \gamma_{\mathbf{k}\nu\nu'} A_{\mathbf{k} \nu'\nu'' }(\omega=0) A_{\mathbf{k} \nu'''\nu}(\omega=0)$ in a plane $k_z=0$. Each band index combination $\nu'',\nu'''$ (for the six spin-orbital bands) is shown in a separate panel. The scale is kept the same within the same channel for better comparison. The panels in the upper triangle correspond to the \BOg channel, and the one in the lower triangle to the \BTg channel. The values are in atomic units.}
\label{fig:vertexcorr_B1gB2g}
\end{center}
\end{figure}

\providecommand{\noopsort}[1]{}\providecommand{\singleletter}[1]{#1}%

\end{document}